\documentclass[12pt, draftclsnofoot, onecolumn]{IEEEtran}
\IEEEoverridecommandlockouts
\usepackage{cite}
\usepackage{amsmath,amssymb,amsfonts}
\usepackage{algorithmic}
\usepackage{graphicx}
\usepackage{textcomp}
\usepackage{xcolor}
\usepackage{algorithm}
\def\BibTeX{{\rm B\kern-.05em{\sc i\kern-.025em b}\kern-.08em
    T\kern-.1667em\lower.7ex\hbox{E}\kern-.125emX}}

\usepackage{xcolor}
\usepackage{subfigure}

\begin{document}

\title{Energy-Efficient Cellular-Connected UAV Swarm Control Optimization}

\author{Yang~Su,\ Hui~Zhou,\ Yansha~Deng and Mischa Dohler
	\thanks{Yang Su, Hui Zhou and Yansha Deng are with the Department of Engineering, King’s College London, London, WC2R 2LS, U.K. (email:\{yang.2.su, hui.zhou, yansha.deng\}@kcl.ac.uk). (Corresponding author: Yansha Deng). This paper was presented in part at the 2022 IEEE Global Communications Conference, December 2022 \cite{10001506}. This work was supported by Engineering and Physical Sciences Research Council (EPSRC), U.K., under Grant EP/W004348/1.}
 \thanks{Mischa Dohler is with the Advanced Technology Group, Ericsson Inc., Silicon Valley, US. (email: mischa.dohler@ericsson.com)}
}

\maketitle

\begin{abstract}
Cellular-connected unmanned aerial vehicle (UAV) swarm is a promising solution for diverse applications, including cargo delivery and traffic control. However, it is still challenging to communicate with and control the UAV swarm with high reliability, low latency, and high energy efficiency. In this paper, we propose a two-phase command and control (C\&C) transmission scheme in a cellular-connected UAV swarm network, where the ground base station (GBS) broadcasts the common C\&C message in Phase \uppercase\expandafter{\romannumeral1}. In Phase II, the UAVs that have successfully decoded the C\&C message will relay the message to the rest of UAVs via device-to-device (D2D) communications in either broadcast or unicast mode, under latency and energy constraints. To maximize the number of UAVs that receive the message successfully within the latency and energy constraints, we formulate the problem as a Constrained Markov Decision Process to find the optimal policy. To address this problem, we propose a decentralized constrained graph attention multi-agent Deep-Q-network (DCGA-MADQN) algorithm based on Lagrangian primal-dual policy optimization, where a PID-controller algorithm is utilized to update the Lagrange Multiplier. Simulation results show that our algorithm could maximize the number of UAVs that successfully receive the common C\&C under energy constraints.
\end{abstract}

\begin{IEEEkeywords}
Cellular-connected UAV swarm network, D2D, multi-agent reinforcement learning, graph attention, Constrained Markov Decision Process 
\end{IEEEkeywords}

\section{Introduction}

Cellular-connected UAV (C-UAV) swarm network has been regarded as a promising solution to tackle various complicated tasks, including cargo delivery, aerial imaging, and traffic control\cite{mozaffari2019tutorial,shakhatreh2019unmanned,9417100}. This is because the cellular technology provides the following advantages: 1) due to the almost ubiquitous accessibility of cellular networks, cellular-connected UAVs provide the pilot with beyond line-of-sight control\cite{lin2018sky}; 2) based on the ultra-reliable low latency communications (URLLC) and massive machine type communications (mMTC) services, cellular-connected UAVs achieve significant performance improvements over conventional UAV communication technologies (such as WiFi and Bluetooth) regarding reliability, security, and data throughput\cite{8470897}; and 3) positioning service in the cellular network provides UAVs with an extra dimension of localization.
On the other hand, to overcome the stringent size, weight, and power (SWAP) limitations of a single UAV, a C-UAV swarm network has been proposed to fully exploit the potential of a C-UAV system in complicated tasks via a group of closely coordinated UAVs\cite{8918497}.

Two different architectures, namely infrastructure-based and infrastructure-assisted architectures, have been proposed. In infrastructure-based architecture\cite{campion2018review}, the cellular ground base station (GBS) receives the flight information from all UAVs in the swarm and transmits their coordination data back. Infrastructure-based architecture embraces the benefit of centralized coordination empowered by GBS with typically powerful computing capabilities. However, the lack of scalability in centralized coordination leads to high round-trip air-ground delay, especially when serving a large number of UAVs. Moreover, due to the fluctuated wireless channel and high mobility of UAVs, providing URLLC service to the whole UAV swarm simultaneously for downlink C\&C communication in infrastructure-based architecture is exceptionally  challenging \cite{mei2019uplink}.
To solve the issues in infrastructure-based architecture, infrastructure-assisted UAV-to-UAV (U2U) communication has been proposed for C-UAV swarm network \cite{9115898}. In infrastructure-assisted U2U communication, UAVs are allowed to communicate with each other directly, and the GBS provides the backbone connectivity between the aerial network formed by the UAVs. Such architecture offers advantages in terms of spectrum and energy efficiency, extended cellular coverage, and decreased backhaul requirements. 

To facilitate U2U communication, Device-to-Device (D2D) technical paradigm has been standardized in 3GPP specification \cite{8214255,han2020towards,9647449}. The so-called "Sidelink" (PC5 radio interface) is a D2D technology in 5G standard and supports broadcast mode and unicast mode \cite{lien20203gpp}. The broadcast mode supports one-to-many compared to the unicast mode's one-to-one paradigm. Nevertheless, the unicast mode is more reliable and energy-saving due to having the following special features: 1) there is ACK/NACK mechanism in D2D unicast mode, which can guarantee the stringent reliability requirement \cite{garcia2021tutorial}; 2) the link adaption is supported in the D2D unicast mode, where the Modulation and Coding Scheme (MCS) cannot be dynamically adjusted based on the quality of radio link \cite{nakamura2002adaptive}; 3) power control is supported for D2D unicast mode, the D2D transmitter could regulate its transmit power dynamically based on the pathloss between the D2D receiver\cite{9088326}.

Authors in \cite{han2020towards} considered multiple GBSs broadcasting the C\&C message to the UAV swarm, however, due to the strong interference from other interfering GBSs, a certain amount of UAVs fail to receive the message. Then, the UAVs that have successfully decoded the C\&C message will broadcast the message to the rest of the UAVs via D2D broadcast communication. This work mainly focused on the analytical expression of the reliability performance and only considered one-round D2D communication, which fails to guarantee the high reliability and low latency in C\&C transmission for all UAVs. 
Moreover, this work did not consider the effect of UAVs' high mobility and limited battery capacity.
In \cite{9647449}, the authors investigated the packet delivery ratio (PDR) of a sidelink-assisted multihop U2U communication model for various scheduling parameters, which focused on the performance analysis without considering the interference from GBS and the power limitations of UAVs. To the best of knowledge, there is no research on the optimal D2D modes selection, i.e. D2D execution policy, based on the surrounding environment to maximize the final message coverage within a limited energy supply.

To deal with more complex communication environment and practical formulation, deep reinforcement learning (DRL) emerges as a promising tool to optimize the D2D execution policy, due to that it solely relies on the self learning of the environment interaction, without the need to derive explicit optimization solutions based on a complex mathematical model \cite{8103164}. 
The DRL has been proposed to optimize the energy efficiency \cite{9237143} and content caching in D2D networks \cite{9448092}.
Authors in \cite{9237143} considered the energy efficiency problem in D2D-enabled heterogeneous cellular networks and proposed a single agent DRL-based method to optimize the communication mode selection (D2D mode or cellular mode) and resource allocation to maximize long-term energy efficiency. In \cite{9448092}, a dynamic and time-varying D2D offloading system was studied considering the uncertain and dynamic content requests, mobility, and the constrained cache capacity of nodes, where a single agent DQN-based solution was proposed to solve the content caching optimization problem. 

These work \cite{9237143,9448092} mainly focused on the single-agent DRL method, where its computational complexity and communication cost increase drastically with the number of user equipment. Hence, the single-agent DRL method is insufficient for large-scale networks requiring high scalability. Note that single-agent DRL method is optimized to maximize its own reward function and may not consider the impact of its actions on other agents in cooperative environments, leading to suboptimal behavior. In contrast, multi-agent DRL can learn to cooperate among multiple agents and adapt to changes\cite{zhang2021multi}, resulting in more effective and robust performance in complex and dynamic environments.

To solve the drawbacks mentioned above, in this paper, we develop a decentralized multi-agent deep reinforcement learning architecture adopting Graph Attention network (GAT)  \cite{velivckovic2017graph} structure.
Our contributions are summarized as follows:

\begin{itemize}
\item 
We propose a two-phase downlink C\&C transmission protocol. In phase \uppercase\expandafter{\romannumeral1}, the GBS control center broadcasts the C\&C message to UAV swarm, but part of UAVs fail to receive the message due to strong interference from other GBSs. In phase \uppercase\expandafter{\romannumeral2}, the UAVs that have received the C\&C message successfully in phase \uppercase\expandafter{\romannumeral1} will execute multi-hop D2D communication to transmit the C\&C message to the rest UAVs. We also consider D2D unicast, D2D broadcast, and hybrid D2D transmissions to fully exploit the D2D mode supported in 5G sidelink standard.

\item
To maximize the number of UAVs that receive the C\&C message successfully after D2D phase within energy constraint, we formulate the problem as a Constrained Markov Decision Process (CMDP) problem, and propose a decentralized constrained graph attention multi-agent Deep-Q-network (DCGA-MADQN) algorithm based on Lagrangian primal-dual policy optimization to find the best policy during D2D communication phase. The UAV that has successfully received the C\&C message will act as an agent, executing actions under D2D unicast, D2D broadcast, and D2D hybrid schemes to maximize the number of UAVs decoding message successfully within the latency and energy constraint. Specifically, we utilize GAT structure to exploit the UAV swarm topology, and PID-Controller algorithm to update the Lagrange multiplier.

\item In the experiments, we develop a realistic simulation framework to evaluate the proposed learning framework under three D2D schemes with different energy constraints. The results show that DCGA-MADQN can maximize the number of UAVs that successfully receive the C\&C message within the energy constraint. It is noted that D2D broadcast and hybrid schemes achieve similar number of UAVs that successfully receive the message, but is much higher than that of D2D unicast scheme.

\end{itemize}

The rest of the papers are organized as follows. Section \rm \uppercase\expandafter{\romannumeral2} presents the system model and problem formulation. Section \uppercase\expandafter{\romannumeral3} provides the detail of DCGA-MADQN algorithm. Section \uppercase\expandafter{\romannumeral4} illustrates the simulation results. Finally, Section \uppercase\expandafter{\romannumeral5} concludes the paper.

\section{System model and problem formulation}
\begin{figure*}[h!]
    \centering
    \subfigure[]{
        \includegraphics[scale=0.6]{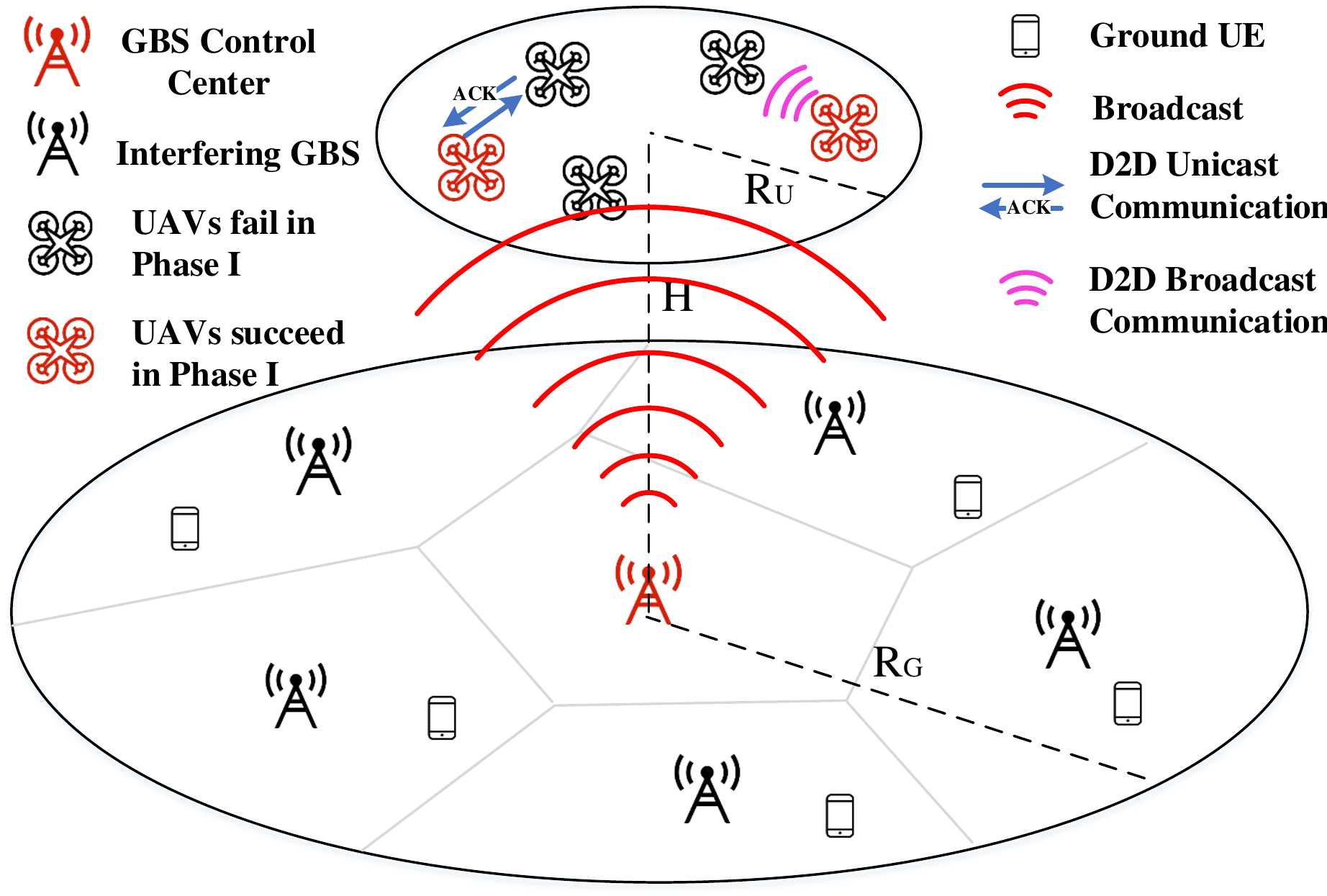}
        \label{fig1}
    }
    \subfigure[]{
	\includegraphics[scale=0.8]{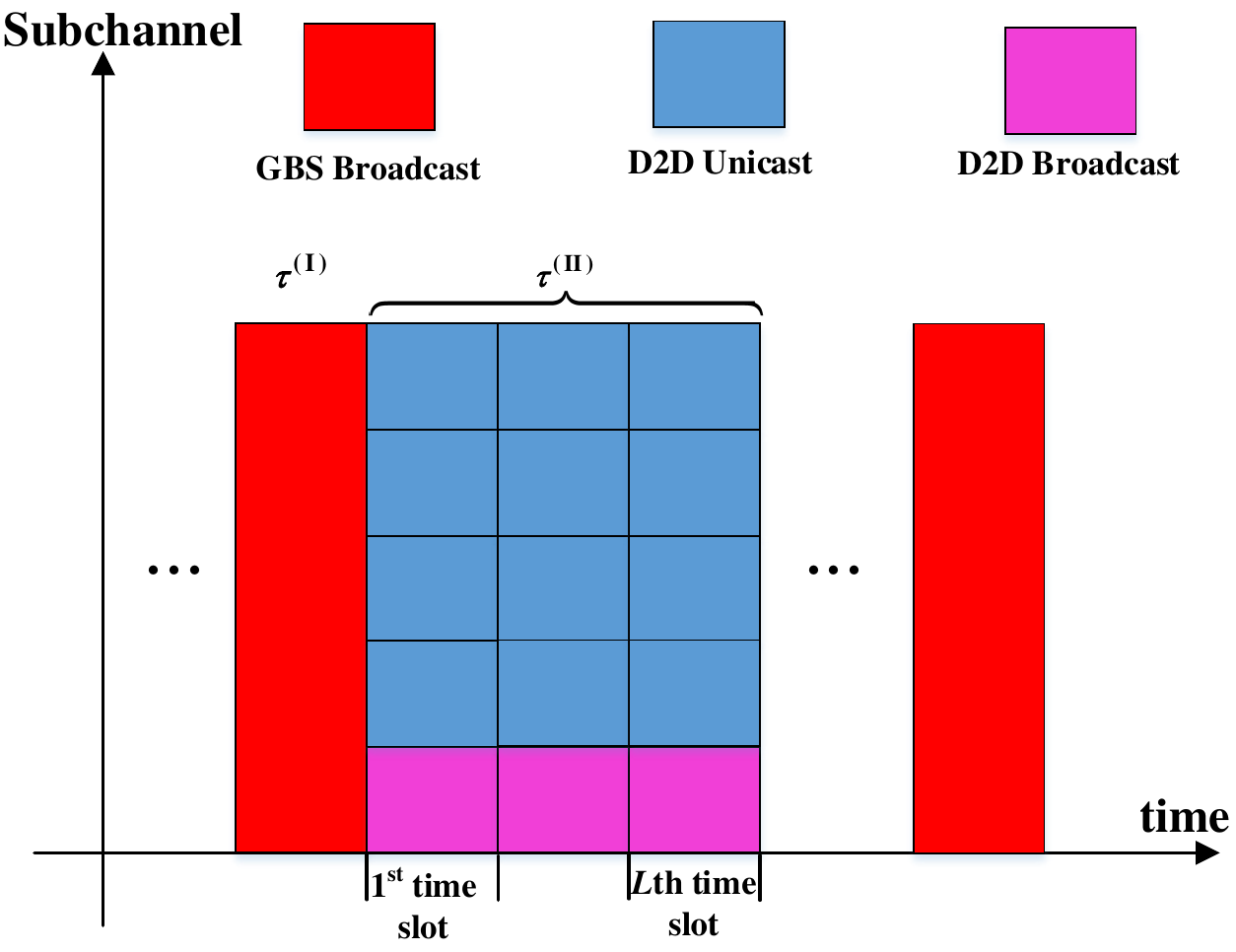}
	\label{}
    }
    \caption{(a) Illustration of a two-phase protocol in celluar connected UAVs system. (b) The time division of two-phase protocol.}
\end{figure*}

As shown in Fig.~\ref{fig1}, we consider the downlink C\&C message transmission in a C-UAV swarm network, where the GBS control center sends the C\&C message to the UAV swarm. It is noted that the interfering GBSs serve the ground user equipments (UEs) using the same frequency band. We assume that each UAV is equipped with a single omnidirectional antenna and each GBS employs \emph{K} antennas, denoted by the antenna indexes set $\mathcal{K} = \{1,2,...,\emph{K}\}$.

The GBS control center will broadcast a common C\&C message to the UAV swarm with the aim to provide flight guidance and cooperation instructions. We assume that there are \emph{N} UAVs in the swarm, denoted by $\mathcal{N} = \{1,2,...,\emph{N}\}$, which are located in a circular horizontal disk with radius \emph{$R_{\rm U}$} and height $\mathrm{H}$. The 3D location of $n_{\rm{th}}$ UAV is denoted as ${u}_n\in \mathbb{R}^{3\times1}$. We also assume that \emph{(M+1)} GBSs are located in a circular ground disk with radius \emph{$R_G$}. Specially, the GBS control center, denoted as $m_0$, is located at the center and just below the center of the UAV swarm, with its 3D location denoted by $\tilde{u}_0\in \mathbb{R}^{3\times1}$. The interfering GBSs are denoted as $\mathcal{M}_1 = \{1,2,...,\emph{M}\}$, and the corresponding 3D location is represented as $\tilde{u}_m\in \mathbb{R}^{3\times1}$.

\subsection{System Model}
Due to the substantial interference from $M$ interfering GBSs, pathloss and channel fading, a certain amount of UAVs with low SINR cannot successfully receive the common C\&C message from the GBS control center. To guarantee the successful transmission of C\&C message to the UAV swarm within the latency requirement $\tau$, we propose a two-phase C\&C transmission protocol. In phase \rm \uppercase\expandafter{\romannumeral1}, the GBS control center will broadcast the common control message to the UAV swarm  within duration $\tau^{(\rm \uppercase\expandafter{\romannumeral1})} < \tau$. In Phase \rm \uppercase\expandafter{\romannumeral2}, the UAVs that have decoded the message successfully will relay the message to other UAVs via multi-hop of D2D communication within remaining duration $\tau^{(\rm \uppercase\expandafter{\romannumeral2})} = \tau - \tau^{(\rm \uppercase\expandafter{\romannumeral1})}$. 

In the following, we will describe the two-phase protocol.

\textit{1) Phase \uppercase\expandafter{\romannumeral1}: Broadcast from GBS control center to UAV swarm}

The GBS control center broadcasts the common control message over a specific frequency band, which is reused by interfering GBSs to serve the ground users. Taking into account the potential line-of-sight (LoS) and non-line-of-sight (NLoS) for flying UAVs, we adopt free-space path loss and Rayleigh fading \cite{9614984,8038869} to model the path loss from the \emph{k}th antenna of \emph{m}th GBS to the \emph{n}th UAV as
\begin{equation} \label{h_mn_k}
h_{m,n}^{k} = \left\{
\begin{array}{lr}
\left(\frac{4\pi d_{m,n}f_{c}}{c}\right)^{\alpha_{\rm \uppercase\expandafter{\romannumeral1}}}\eta_{\mathrm{LoS}}\beta_{m,n}^{k}, \quad P_{\mathrm{LoS}}^{m,n}\\
\left(\frac{4\pi d_{m,n}f_{c}}{c}\right)^{\alpha_{\rm \uppercase\expandafter{\romannumeral1}}}\eta_{\mathrm{NLoS}}\beta_{m,n}^{k}, \quad P_{\mathrm{NLoS}}^{m,n} = 1 - P_{\mathrm{LoS}}^{m,n}\!,\\
\end{array}
\right.
\end{equation}
where $d_{m,n}$ is the distance between the \emph{m}th GBS and the \emph{n}th UAV, $f_{c}$ is the broadcast frequency, $\eta_{\mathrm{LoS}}$ and $\eta_{\mathrm{NLoS}}$ are the path loss coefficients in LoS and NLoS cases, $c$ is the speed light, and $\alpha_{\rm \uppercase\expandafter{\romannumeral1}}$ is the path loss exponent. The $\beta_{m,n}^{k}$ is the small-scale Rayleigh fading from the $k$th antenna of $m$th GBS to the $n$th UAV, which follows $\mathcal{CN}(0,1)$. In (\ref{h_mn_k}), we adopt the LoS probability of the broadcast transmission as\cite{7510820,al2014optimal,7412759}
\begin{equation} \label{P_LoS_mn}
P_{\mathrm{LoS}}^{m,n} = \frac{1}{1+a\ {\rm exp}(-b(\theta_{m,n} - a))},
\end{equation}
where $\theta_{m,n} = \frac{180}{\pi} \times arcsin\left(\frac{\mathrm{H}}{d_{m,n}}\right)$ is the elevation angle of the $n$th UAV, $H$ is the height of the UAV, $a$ and $b$ are positive constants that depend on the environment.

Hence, based on (\ref{h_mn_k}) and (\ref{P_LoS_mn}), the channel from the $k$th antenna of $m$th GBS to $n$th UAV can be obtained as
\begin{equation} \label{h_mn_k1}
h_{m,n}^{k} = (P_{\mathrm{LoS}}^{m,n}\eta_{\mathrm{LoS}}+P_{\mathrm{NLoS}}^{m,n}\eta_{\mathrm{NLoS}})\left(\frac{4\pi d_{m,n}f_{c}}{c}\right)^{\alpha_{\rm \uppercase\expandafter{\romannumeral1}}}\beta_{m,n}^{k}.
\end{equation}

The transmitted signal of the GBS control center $m_0$ can be presented as
\begin{equation}
x_0^{(\rm \uppercase\expandafter{\romannumeral1})}=\sqrt{P}s_0,
\end{equation}
where $P$ is the identical maximal transimit power of GBSs, $s_0$ is the common control message with $\mathbb{E}\{|s_0|^2\}=1$.

The transmitted signal of interfering GBS $m$ is given by
\begin{equation}
x_m^{(\rm \uppercase\expandafter{\romannumeral1})}=\sqrt{P}s_m,\quad m\in \mathcal{M}_1,
\end{equation}
where $s_m$ is the transmitted message for serving ground user equipments, with $\mathbb{E}\{|s_m|^2\}=1$.

The received signal of the $n$th UAV in Phase \rm \uppercase\expandafter{\romannumeral1} can be written as
\begin{align}
y_n^{(\rm \uppercase\expandafter{\romannumeral1})}
&=\sqrt{P}\sum\limits_{k\in \mathcal{K}}h_{m_0,n}^{k}s_{0}+\sqrt{P}\underset{m\in \mathcal{M}_1,k\in \mathcal{K}}{\sum{\sum}}{h_{m,n}^{k}s_{m}}\notag \\&\quad+z_n^{(\rm \uppercase\expandafter{\romannumeral1})},\quad n\in \mathcal{N}.
\end{align}
where $z_n^{(\rm \uppercase\expandafter{\romannumeral1})}\sim\mathcal{CN}(0,\sigma ^2)$ is the additive white Gaussian noise at the $n$th UAV.

As a result, the SINR of the received signal at $n$th UAV is
\begin{equation}
{\rm SINR}_n^{(\rm \uppercase\expandafter{\romannumeral1})}=\frac{P\left|\sum\nolimits_{k\in \mathcal{K}}h_{m_0,n}^{k}\right|^2}{\sum\nolimits_{m\in \mathcal{M}_1}P\left|\sum\nolimits_{k\in \mathcal{K}}h_{m,n}^{k}\right|^2 + \sigma ^2},\quad n\in\mathcal{N}.
\end{equation}

It is noted that the $n{\rm th}$ UAV can successfully decode 
the common C\&C if the SINR is higher than the threshold $\gamma_{\rm \uppercase\expandafter{\romannumeral1}}$. Therefore, the $N$ UAVs can be divided into two groups $\mathcal{N}_s$ and $\mathcal{N}_f$.
\begin{equation}
\left\{
\begin{array}{lr}
\mathcal{N}_s=\{n\mid n\in\mathcal{N}, {\rm SINR}_n^{(\rm \uppercase\expandafter{\romannumeral1})}\geq\gamma_{\rm \uppercase\expandafter{\romannumeral1}}\}\\
\mathcal{N}_f= \mathcal{N} \setminus \mathcal{N}_s
\end{array}
\right.,
\end{equation}
where $\mathcal{N}_s$ represents the UAVs that have successfully decoded the broadcasted common C\&C message, and has a total number of $N_s$ UAVs; and $\mathcal{N}_f$ represents the UAVs that fail to decode the broadcasted common C\&C message, and has a total number of $N_f$ UAVs. Since $\mathcal{N}_s$ and $\mathcal{N}_f$ represent
the initial UAVs set before the D2D unicast transmission, we also refer them to $\mathcal{N}_s^0$ and $\mathcal{N}_f^0$ , respectively.
% $N_s={n, SINR_n>=\gamma}$

\textit{2) Phase \uppercase\expandafter{\romannumeral2}: D2D communication among UAVs}

After the broadcast of the GBS control center, the initial UAVs $\mathcal{N}_s^0$ which have received the common control message successfully will transmit the message to the rest of UAVs $\mathcal{N}_f^0$ via multiple cycles of D2D communication within latency constraint. The duration of Phase \rm \uppercase\expandafter{\romannumeral2} $\tau^{(\rm \uppercase\expandafter{\romannumeral2})}$ could contain $L$ time slots, where each time slot occupies $\Delta t$. Hence, the cumulative duration of all time slots is $t_{\rm slots} = L\Delta t$. In each time slot, the successful UAV will execute one cycle of D2D communication, which includes two possible operation modes: unicast mode and broadcast mode.
We define $\mu_{i,\rm{u}}^{t}$ and $\mu_{i,\rm{b}}^{t}$ as the operation mode indicator of the $i$th  
UAV $(i \in \mathcal{N}_s^{t-1})$ in $t$th time slot,
\begin{equation}
\mu_{i,\rm{u}}^{t}= \begin{cases}1, & \text {unicast mode}
\\ 0, & \text {broadcast or idle mode}\end{cases},
\end{equation}

\begin{equation}
\mu_{i,\rm{b}}^{t}= \begin{cases}1, & \text {broadcast mode}
\\ 0, & \text {unicast or idle mode}\end{cases},
\end{equation}

\begin{equation}
\mu_{i,\rm{u}}^{t} + \mu_{i,\rm{b}}^{t} <=1.
\end{equation}

Depending on the selection of operation mode, the UAVs in the set of $ \mathcal{N}_s^{t-1}$ could be further divided into three groups:
\begin{equation}
\left\{
\begin{array}{lr}
\Theta_{\rm u}^t=\{n\mid n\in\mathcal{N}_s^{t-1}, \mu_{n,\rm{u}}^{t} = 1 \}\\
\Theta_{\rm b}^t=\{n\mid n\in\mathcal{N}_s^{t-1}, \mu_{n,\rm{b}}^{t} = 1\}\\
\Theta_{\mathrm{idle}}^t=\{n\mid n\in\mathcal{N}_s^{t-1}, \mu_{n,\rm{u}}^{t} = 0\; \& \; \mu_{n,\rm{b}}^{t} = 0 \}
\end{array}
\right.,
\end{equation}
where $\Theta_{\rm u}^t$ represents the UAVs working at unicast mode in $t$th time slot, $\Theta_{\rm b}^t$ represents the UAVs working at broadcast mode in $t$th time slot and
$\Theta_\mathrm{idle}^t$ represents the UAVs that remain idle in $t$th time slot.
It is noted that when the $i$th UAV in set $\Theta_{\rm u}^t$ chooses the $n$th UAV from the set $\mathcal{N}_f^{t-1}$ as the D2D unicast receiver and connects successfully, then the target $n$th UAV will be removed from $\mathcal{N}_f^{t-1}$, and added to $\mathcal{N}_s^{t-1}$, which becomes part of sets $\mathcal{N}_s^{t}$ and $\mathcal{N}_f^{t}$ in $t$th time slot.

Under D2D unicast mode, the $i$th UAV utilizes the power control algorithm to compensate the path loss following \cite{9839227,8408843}, which is shown as
\begin{equation}
\label{unicast transmit power}
\tilde{P}_{i,\rm{u}}= \mathrm{min}\{\xi d_{i,n}^{\mathrm{\alpha_{I I}}}, \tilde{P}_{\rm{max}}\}, \quad i \in \Theta_{\rm{u}}^t, n \in \mathcal{N}_{\rm{s}}^{t-1},
\end{equation}
where $\xi$ is the power parameter, $d_{i,n}$ is the distance between the D2D unicast pairs, $\mathrm{\alpha_{I I}}$ is the pathloss exponent and $\tilde{P}_{\rm{max}}$ is the maximum transmit power of UAV.

The received signal of the $n$th UAV from the $i$th UAV can be derived as
\begin{equation}
y_{n,u}^{(\rm \uppercase\expandafter{\romannumeral2})}=\sqrt{\tilde{P}_{i,u}}h_{i,n}s_{0} + z_n^{(\rm \uppercase\expandafter{\romannumeral2})},
\end{equation}
where the path loss $h_{i,n} = d_{i,n}^{-\alpha_{\rm \uppercase\expandafter{\romannumeral2}}}\beta_{i,n}$, and $\beta_{i,n}$ is the Rayleigh fading following $\mathcal{CN}(0,1)$. The $z_n^{(\rm \uppercase\expandafter{\romannumeral2})}\sim\mathcal{CN}(0,\sigma ^2)$ is the additive white Gaussian noise at the $n$th UAV.

Therefore, we derive the received SINR at the $n$th UAV as
\begin{equation}\label{SINR_n2}
{\rm SINR}_{n,\rm{u}}^{(\rm \uppercase\expandafter{\romannumeral2})}=\frac{\tilde{P}_{i,\rm{u}}\left|h_{i,n}\right|^2}{\sum\nolimits_{m\in \mathcal{M}_1}P\left|\sum\nolimits_{k\in \mathcal{K}}h_{m,n}^{k}\right|^2+\sigma ^2}.
\end{equation}
When ${\rm SINR}_{n,\rm{u}}^{(\rm \uppercase\expandafter{\romannumeral2})}$ is above the threshold $\gamma_{\rm \uppercase\expandafter{\romannumeral2}}$, the D2D unicast transmission between UAV $i$ and UAV $n$ is identified as successful, and the $n$th UAV will send back an ACK signal to UAV $i$ to indicate the success of transmission.

The total energy consumed by the D2D unicast communication in time slot $t$ is calculated as,
\begin{equation}
E_{i,\rm{u}}=\left(\kappa \tilde{P}_{i,\rm{u}}+\tilde{P}_{i,\rm{o}}\right) \Delta t, \quad i \in \Theta_{\rm{u}}^t,
\end{equation}
where $\kappa$ is the conversion factor of the power amplifier from electric power to RF power and $\tilde{P}_{i,o}$ is the electronic power consumption overhead incurred in the communication module to encode the common control message \cite{zhao2016fundamental}.

We consider that all the UAVs in $\Theta_b^t$ will broadcast the signal at the same subchannel, and the interior clock is synchronized through GNSS. In D2D broadcast mode, the maximum transmit power is utilized to enlarge the coverage as

\begin{equation}
\label{broadcast transmit power}
\tilde{P}_{i,\rm{b}}= \tilde{P}_{\rm{max}}, \quad i \in \Theta_{\rm{b}}^{t}, n \in \mathcal{N}_{\rm{s}}^{t-1}.
\end{equation}

The received signal of the $n$th UAV can be derived as
\begin{equation}
y_{n,{\rm b}}^{(\rm \uppercase\expandafter{\romannumeral2})}=\sum\limits_{i \in \Theta_{\rm b}^t} \sqrt{\tilde{P}_{i,{\rm b}}} h_{i,n}s_{0} + z_n^{(\rm \uppercase\expandafter{\romannumeral2})},
\end{equation}

The corresponding SINR of the $n$th UAV is

\begin{equation}
\mathrm{SINR}_{n,\rm{b}}^{(\rm \uppercase\expandafter{\romannumeral2})}=\frac{\tilde{P}_{i,\rm{b}}\left|\sum\nolimits_{i \in \Theta_{\rm{b}}^t}h_{i,n}\right|^2}{\sum\nolimits_{m\in \mathcal{M}_1}P\left|\sum\nolimits_{k\in \mathcal{K}}h_{m,n}^{k}\right|^2+\sigma ^2}.
\end{equation}
When ${\rm SINR}_{n,{\rm b}}^{(\rm \uppercase\expandafter{\romannumeral2})}$ is above the threshold $\gamma_{\rm \uppercase\expandafter{\romannumeral2}}$, UAV $n$ will receive D2D broadcast signal successfully.

The total energy consumed by the D2D broadcast communication in time slot $t$ is calculated as,
\begin{equation}
E_{i,\rm{b}}=\left(\kappa \tilde{P}_{i,\rm{b}}+\tilde{P}_{i,\rm{o}}\right) \Delta t, \quad i \in \Theta_{\rm{b}}^t.
\end{equation}

Apart from Transmitting state (TX), i.e., unicast mode or broadcast mode mentioned above, each UAV can also be in Receiving state (RX) or idle state. The corresponding energy consumption is given by
\begin{equation}
E_{i,\rm{r}}=\tilde{P}_{i,\rm{r}}\Delta t, \quad i \in \mathcal{N},
\end{equation}
\begin{equation}
E_{i,\rm{idle}}=\tilde{P}_{i,\rm{idle}}\Delta t,  \quad i \in \Theta_{\rm{idle}}^t,
\end{equation}
where $\tilde{P}_{i,\rm{r}}$ and $\tilde{P}_{i,\rm{idle}}$ are constant power of receiving state and idle state.
Hence, the energy consumption of UAV $i$  in time slot $t$ during Phase \uppercase\expandafter{\romannumeral2} is calculated as,
% \begin{equation}
% E_{i,total}=E_{i,b}n_b + E_{i,u}n_u + E_{i,r}n_r + E_{i,idle}n_{idle},
% \end{equation}

\begin{equation}
E_{i}(t)= \begin{cases}\mu_{i,\rm{u}}^{t}E_{i,\rm{u}}+\mu_{i,\rm{b}}^{t}E_{i,\rm{b}}, & \text {transmitting state}
\\ E_{i,\rm{r}}, & \text {receiving state}
\\ E_{i,\rm{idle}}, & \text {idle state} \end{cases}.
\end{equation}

\subsection{Mobility Model}
The widely applied random waypoint (RWP) mobility model \cite{hyytia2007random} is adopted to model each UAV's mobility. Each node starts by pausing for a fixed number of seconds, called the pause period. When the pause period is elapsed, the node chooses a random end position within the area of simulation and moves towards the end position with a randomly chosen speed. Upon arrival at the end position, it stops and waits for a moment before starting its journey to a newly chosen end position. This procedure is repeated until the simulation period is elapsed. Due to the stringent latency requirement, we assume the UAVs are static during
the execution of Phase \uppercase\expandafter{\romannumeral1} and Phase \uppercase\expandafter{\romannumeral2}, which include GBS broadcasting, and UAV D2D unicast.

\subsection{Problem Formulation}

In this paper, we consider three D2D communication schemes: Unicast, Broadcast, and Hybrid schemes. In Unicast scheme, each UAV can only perform D2D unicast transmission or remain idle in each time slot. In Broadcast scheme, each UAV can only operate in D2D broadcast mode or remain idle during each cycle. For hybrid scheme, each UAV can choose D2D unicast transmission, D2D broadcast transmission or keep idle during each round.

We focus on the downlink C\&C transmission from the GBS control center to the UAV swarm in the mission area with the aim to maximize the number of UAVs
that successfully receive the common C\&C message under the overall energy consumption constraint and latency constraint $\tau$. The problem can be formulated as 
\begin{equation}\label{optimization}
\begin{aligned}
\max \limits_{\pi (A^{t} | O^{t})} \quad & \sum_{k=t}^{\infty}\gamma^{k-t} \mathbb{E}_{\pi}[N_{s}^{k}]\\
\textrm{s.t.} \quad &C^{k}\leq \rm{E}_{\rm{c}},
\end{aligned}
\end{equation}
where $\pi$ is the policy that maps the current observation $O^{t}$ to the probabilities of actions, $\gamma \in [0,1)$ is the discount factor for the performance in future slots, $\rm{E}_{\rm{c}}$ is the overall energy constraint, $C^k$ is the overall energy cost of all UAVs from time slot $0$ to time slot $k$ and represented as
\begin{equation}
C^{k} = \sum_{h=0}^{k}\sum_{i=1}^{\mathrm{N}}E_{i}\left(h\right).
\end{equation}

The problem in (\ref{optimization}) is a constrained Markov Decision Process (C-MDP) problem\cite{altman1999constrained} and can be transformed into the following unconstrained form based on the Lagrangian primal-dual policy optimization technique,
\begin{equation}\label{cmdp}
\min _{\lambda \geq 0} \max _{\pi} \sum_{k=t}^{\infty}\gamma^{k-t} \mathbb{E}_{\pi}[N_s^{k}]-\lambda \left({\rm E_c}-C^k\right),
\end{equation}
where $\lambda$ is the lagrange multiplier, and $\pi$ is the policy.

\begin{figure}[htbp]
\centerline{\includegraphics[scale=0.6]{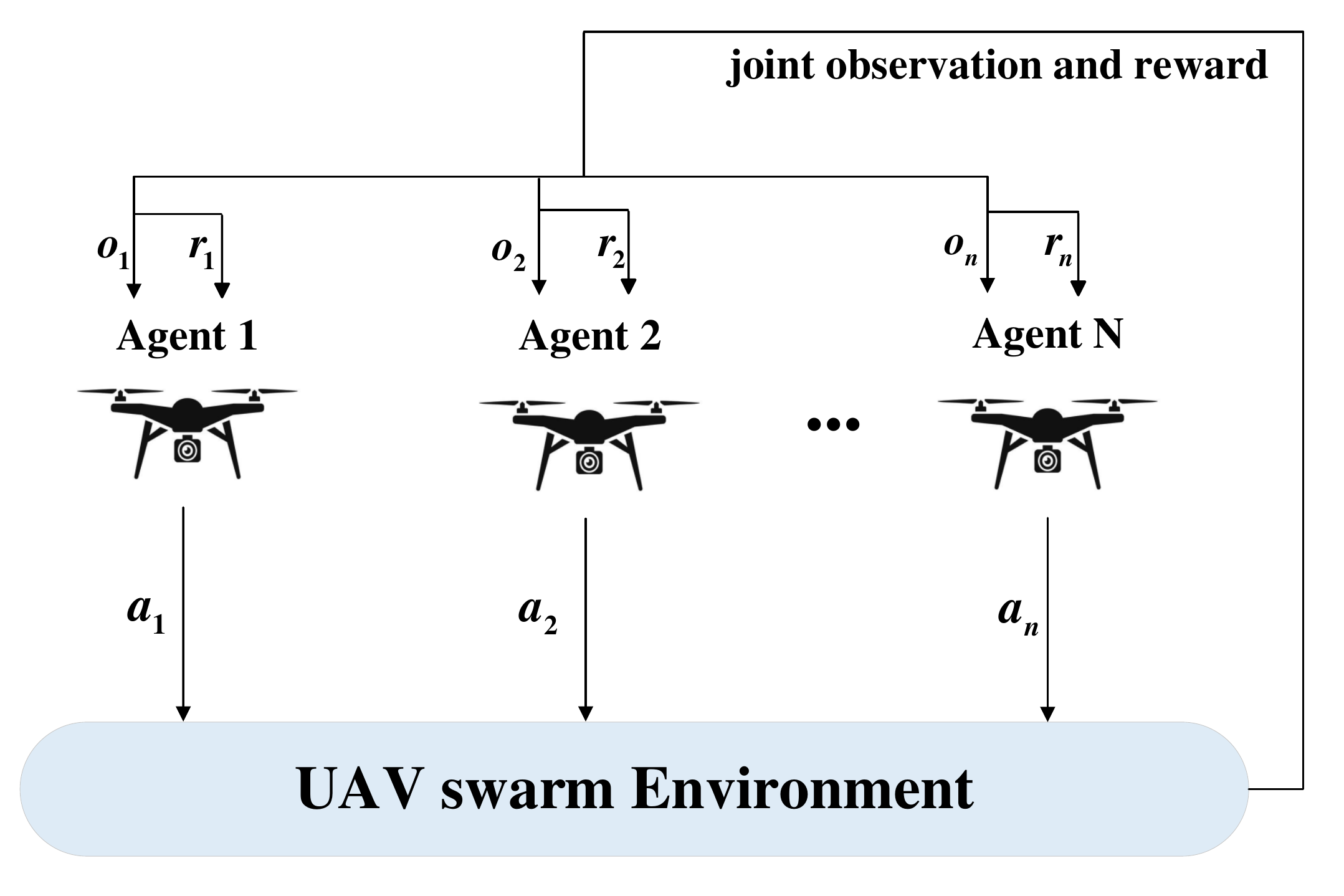}}
\caption{The framework of Multi-agent deep reinforcement learning.}
\label{The framework of Multi-agent deep reinforcement learning}
\end{figure}
 
\section{Multi agent reinforcement learning based on Graph attention network}
In this section, we will introduce a decentralized constrained multi-agent deep reinforcement learning algorithm to solve the optimization problem in (\ref{cmdp}). The architecture of the multi-agent deep reinforcement learning is shown in  Fig.~\ref{The framework of Multi-agent deep reinforcement learning}, during each time slot of the D2D communication phase, the UAVs that fail to receive the message will not have any action and will wait for other UAVs' D2D connection. Once the UAVs receive the message successfully, they will act as agents, acquire the observation from the environment and  execute different actions based on different D2D communication schemes.

\textbf{Observation:}
For all of the three schemes, we assume that each agent has a global view of the entire operational environment, including the locations, message receiving statuses and energy consumption of all other agents. The local observation of agent $i$ encompasses the aforementioned information and is represented as $O_i^t = ({\mu}_i^t, {\varphi}_i^t, e_i^t)$. Specifically, we denote ${\mu}_i^t$ as all UAVs' 3D locations with ${\mu}_i^t = \{u_1, u_2, ... , u_n\}, n\in\mathcal{N}$. The message receiving status of all UAVs from the view of agent $i$ will be represented as ${\varphi}_i^t = \{{\varphi}_1, {\varphi}_2, ... , {\varphi}_n\}, n\in\mathcal{N}$, where
\begin{equation} \label{varphi_n}
{\varphi}_n = \left\{
\begin{array}{lr}
0, \quad {\rm failure}\\
1, \quad {\rm success}\\
\end{array}
\right..
\end{equation}
The $e_i^t$ is utilized to represent the value difference between the overall energy constraint and the energy consumed by UAVs until $t$th time slot, where
\begin{equation} 
e_i^t = {\rm E_c}-C^{t-1}.
\end{equation}

\textbf{Action:}
The action space for the three D2D communication schemes are presented in detail as follows.

1) Unicast scheme: During each time slot of phase \rm \uppercase\expandafter{\romannumeral2}, agent $i$ will choose one UAV from the remaining $N-1$ UAVs as the target to connect, or it will take no action and remain idle in this time slot. The size of action space is $N$, and the action of agent $i$ in $t$th time slot is expressed as 
\begin{equation}
A_i^t = \left\{
{\rm unicast}, \  {\rm idle}\\
\right\}.
\end{equation}

2) Broadcast scheme: agent $i$ is required to make a decision about whether executing broadcast in each time slot of phase \rm \uppercase\expandafter{\romannumeral2}, therefore, its action space size is $2$. During $t$th time slot, the action of agent $i$ can be denoted as
\begin{equation}
A_i^t = \left\{
{\rm broadcast}, \  {\rm idle}
\right\}
\end{equation}

3) Hybrid scheme: During each time slot of phase \rm \uppercase\expandafter{\romannumeral2}, agent $i$ needs to first choose the operating mode: D2D broadcast mode, D2D unicast mode or idle mode. If agent $i$ chooses to work at unicast mode, it needs to make decisions further, which means select one UAV from the remaining $N-1$ UAVs as the target to connect. Hence, the size of action space is $N+1$. 
The action of agent $i$ in $t$th time slot is written as 
\begin{equation}
A_i^t =  \{ {\rm unicast},\ {\rm broadcast},\ {\rm idle}\}.
\end{equation}

\textbf{Reward:}
We consider three D2D communication schemes adopting the same reward design. As our goal is to maximize the number of UAVs that successfully receive the C\&C message under latency constraints, the reward function should be defined according to the additional number of UAVs that receive the message successfully in the current $t$th time slot compared to the $(t-1)$th time slot, which is expressed as

\begin{equation}
{R}_i^t = N_{add}^t, \quad (i \in \mathcal{N}_s^{t-1}).
\end{equation}

\begin{algorithm}
\caption{Constrained Multi-Agent Deep-Q-network}
\label{MADQN}
\begin{algorithmic}[1]
\STATE Initialization:
\FOR{all $i \in N$}
\STATE Initialize replay memory $D_i$ to capacity $N_D$, current Q-network $Q_{i}(O_i,A_i;\theta_i)$, target Q-network $\hat{Q}_{i}(O_i^{\prime},A_i^{\prime};\hat{\theta_i})$, Lagrange Multiplier $\lambda _i$\
\ENDFOR
\WHILE {not\ at\ max\_episode}
\FOR{each agent $i \in\mathcal{N}_s$}
\STATE Achieve local observation $O_i$
\IF{with probability $1-\epsilon$} 
\STATE $A_i = max_{A}Q_i(O_i,A_i;\theta_i)$
\ELSE
\STATE Select random action $A_i$
\ENDIF
\STATE Update the reward $R_i$
\STATE Achieve the next local observation $O_i^{\prime}$
\STATE Store transition $(O_i,A_i,R_i,O_i^{\prime},c_i)$ in the replay memory $D_i$
\STATE Sample a mini-batch of $M$ transitions from $D_i$
\IF{$O_i^{\prime}$ is terminal} 
\STATE $y_i = R_i$
\ELSE
\STATE $y_i = R_i +  \gamma max_{A_{i}^{\prime}}\hat{Q}_{i}(O_i^{\prime},A_i^{\prime};\hat{\theta_i})-\lambda_i c_{i}$
\ENDIF
\STATE Using stochastic gradient to minimize the loss
\STATE $L = (y_i-Q_{i}(O_i,A_i;\theta_i))^2$
\STATE Update the target Q-network $\hat{Q}_{i}$ and Lagrange Multiplier $\lambda _i$
\ENDFOR
\ENDWHILE
\end{algorithmic}
\end{algorithm}

\subsection{Constrained Multi-Agent Deep Q Network}
We adopt a decentralized constrained multi-agent deep Q-network to solve the optimization problem, where each agent is trained based on its own local observation. Each agent $i$ has its own Q-network, target Q-network, and replay memory $D_i$. For different agents, the architecture of their own Q-network and target Q-network is the same. However, they will not share the parameters considering the stringent time requirement and communication cost in a wireless environment. We use the $\epsilon-greedy$ algorithm to balance the exploration and exploitation; the agent chooses the optimal target with a high probability $1-\epsilon$ or selects a random target with probability $\epsilon$. For each agent $i$, it samples a random minibatch of $M$ samples from replay memory $D_i$ and uses a stochastic gradient to minimize the Q-loss. The replay memory could make more efficient use of the experiences during the training, prevent the forgetting of previous experiences and diminish the correlation between experiences.
The weights of the target Q-network $\hat{Q}_{i}$ will be updated by slowly track the learned  Q-network ${Q}_{i}$ :
\begin{equation}
\hat{\theta_i} = \beta\theta_i + (1-\beta)\hat{\theta_i},
\end{equation}
where the $\beta$ is an interpolation parameter and much less than 1. This slow update method could greatly improve the stability of learning.

The state action value for each UAV $i$ is calculated as
\begin{equation}
Q_{i}\left(O_{i}, A_{i}\right)=R_{i}+\gamma \max _{A_{i}^{\prime}} Q_{i}\left(O_{i}^{\prime}, A_{i}^{\prime}\right)-\lambda_i c_{i}\left(O_{i}, A_{i}\right)
\end{equation}
where $O_{i}$ is current state, $A_{i}$ is current action, $R_{i}$ is reward, $O_{i}^{\prime}$ is next state, $A_{i}^{\prime}$ is next action, $\lambda_{i}$ is Lagrange multiplier, $c_{i}$ is energy cost of current action, $\gamma$ is the discount factor, which determines the balance between the current state-action value and future state-action values.

In this study, we adopt a control-theoretic approach to update the Lagrange multiplier $\lambda$ in the learning algorithm\cite{stooke2020responsive}. This is achieved by interpreting the overall learning process as a dynamical system and utilizing the Proportional-Integral-Derivative (PID) control rule. The implementation details are outlined in Algorithm \ref{PID}.

The cumulative energy cost of all UAVs, as perceived by UAV $i$, in the last time slot of the D2D phase II is represented by $C_i^L$. The PID update rule allows for fine-tuned control of the Lagrange multiplier $\lambda$ by considering three key components: proportional control, integral control, and derivative control. The proportional control term $K_P \Delta_i$ is proportional to the difference value $\Delta_i$ between the current energy cost and the energy constraint. This term accelerates the response to constraint breaches and reduces oscillations in the system. The integral control term $K_I I_{i}$ considers the accumulated past values of $\Delta_i$, and the derivative control term $K_D D_i$ is an estimate of the future trend of $\Delta_i$. This term acts against increasing energy costs while allowing for decreases, projected as (.)+.

The utilization of the PID control rule in this study is motivated by the need for precise control of the Lagrange multiplier $\lambda$ in the learning algorithm. 
The Lagrange multiplier $\lambda$ acts as a weight that balances the trade-off between maximizing the system's objective function and satisfying the energy constraint. Specifically, it penalizes the objective function when the energy consumption of the system exceeds the specified energy budget. The penalty is proportional to the amount by which the energy consumption exceeds the energy budget. 
By adjusting the value of $\lambda$, the learning algorithm can control the energy consumption of the system. A higher value of $\lambda$ results in a larger penalty on the objective function for exceeding the energy constraint, thereby enforcing a stricter energy constraint. In contrast, a lower value of $\lambda$ results in a smaller penalty and a more relaxed energy constraint.
The PID control rule provides a physical interpretation of the learning process, allowing for fine-tuned control of the Lagrange multiplier $\lambda$ and enabling optimization of the overall system performance.

\begin{algorithm}
\caption{PID-Controller Lagrange Multiplier}
\label{PID}
\begin{algorithmic}[1]
\STATE Select tuning parameters: $K_P, K_I, K_D \geq 0$
\FOR{all $i \in N$}
\STATE Integral: $I_{i} \leftarrow 0$
\STATE Previous Cost $C_{i,prev} \leftarrow 0$
\ENDFOR
\WHILE {$not\ at\ max\_episode$}
\FOR{each agent $i$}
\STATE Receive cost $C_{i}^{L}$
\STATE $\Delta_i \leftarrow C_{i}^{L} - \rm{E}_{\rm{c}}$
\STATE $D_i \leftarrow \left(C_{i}^{L} - C_{i,prev}^{L}\right)_{+}$
\STATE $I_i \leftarrow \left(I_{i} + \Delta_i\right)_{+}$
\STATE $\lambda_i \leftarrow \left(K_P \Delta_i + K_I I_{i} + K_D D_i\right)_{+}$
\STATE $C_{i,prev}^{L} \leftarrow C_{i}^{L}$
\ENDFOR
\ENDWHILE
\end{algorithmic}
\end{algorithm}

\subsection{Graph Attention Network Architecture}

\begin{figure*}[h]
\centerline{\includegraphics[scale=0.58]{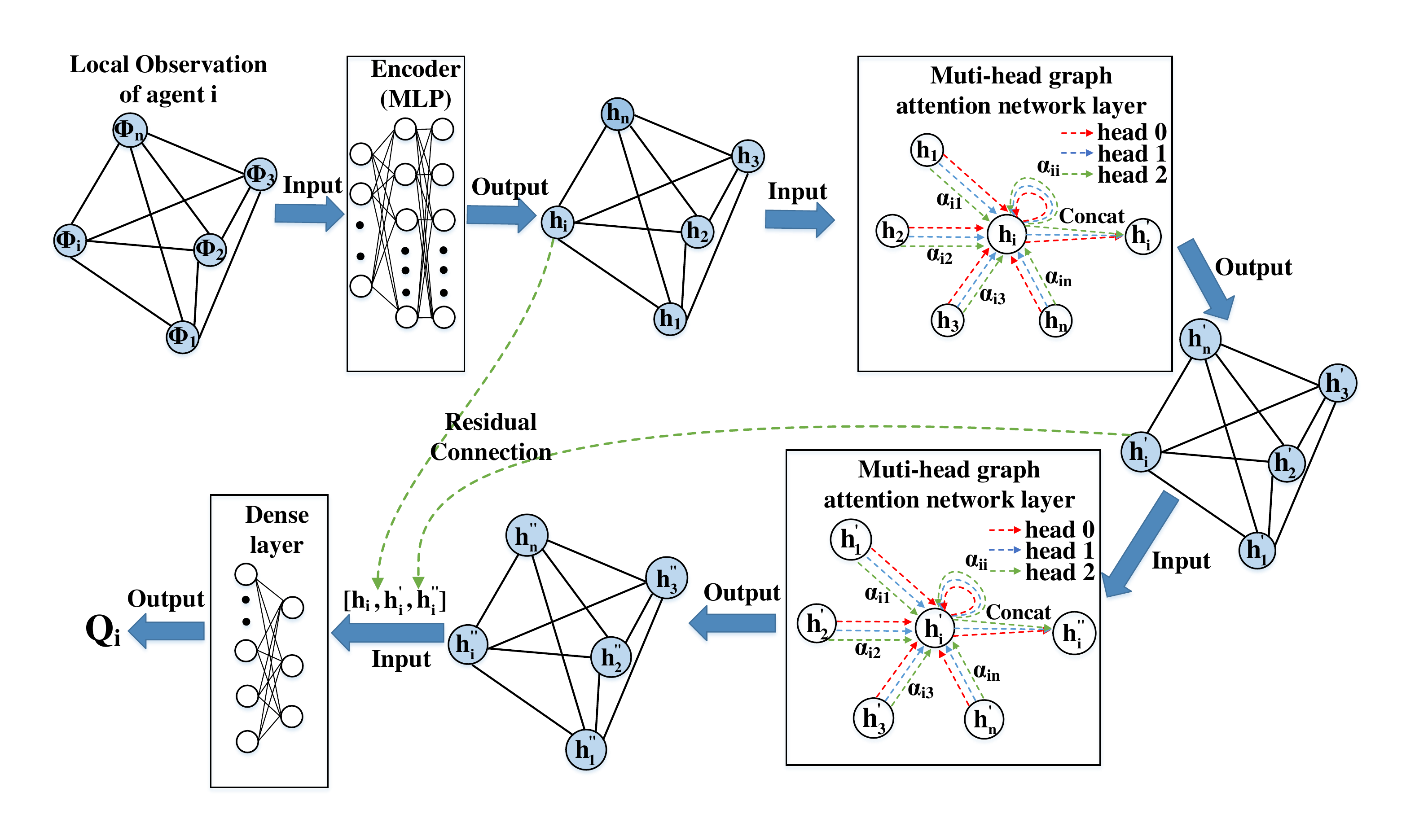}}
\caption{The architecture of neural network in DQN of agent $i$.}
\label{GAT_architecture}
\end{figure*}

The local observation of each agent is represented as a fully-connected graph, where different nodes represent different UAVs containing corresponding raw information, and there exists an undirected edge between any two nodes.

We employ a graph attention network to approximate the Q-values. The observation $O_i$ of agent $i$ can be decomposed into ${\phi_1, \phi_2, ..., \phi_n}, n\in\mathcal{N}$, where $\phi_n\ = \{u_n, \varphi_n, e_i\}$. The $\phi_n$ corresponds to the raw information of node $n$ in the graph, where $u_n$ represents the 3D location of node $n$, $\varphi_n$ denotes the message receiving status of node $n$, and $e_i$ represents the difference between the overall energy constraint and the energy consumed by the UAVs.

In order to expedite the convergence of the neural network model during training, we normalize the observation data by scaling. Normalization is often employed in machine learning to facilitate the optimization process by ensuring that the input data is consistent and comparable across different features. In our case, as the UAVs are constrained to move in a circular region with a radius of $R_U$ and height $H$, we scale the 3D location data $u_n$ by dividing it by $R_U$ and $H$. This ensures that the scale of the input features is similar, which can improve the convergence speed and increase the model's stability.

\begin{equation}
\label{u_n_prime}
u_n=\left\{x_n,y_n,z_n\right\}\rightarrow u_n^{\prime}=\left\{\frac{x_n}{R_U},\frac{y_n}{R_U},\frac{z_n}{H}\right\}
\end{equation}
When a UAV implements D2D communication, broadcast mode incurs the highest energy cost because it utilizes its maximum transmit power. Hence, for the energy difference value $e_i$, we adopt $E_{i,b}$ as the scale factor.
\begin{equation}
\label{e_i_prime}
e_i \rightarrow e_i^{\prime}= \frac{e_i}{E_{i,b}}
\end{equation}
Then, we could obtain newly normalized data $\phi_n^{\prime}\ = \{n, u_n^{\prime}, \varphi_n, e_i^{\prime}\}$, where $u_n^{\prime}$ is given in (\ref{u_n_prime}), $e_i^{\prime}$ is given in (\ref{e_i_prime}).

As shown in Fig.~\ref{GAT_architecture}, the architecture of the neural network consists of four parts: one encoder layer, two multi-head graph attention network layer and one dense layer. The node information $\phi_i^{\prime}$ will be encoded into a feature vector $h_i$ by Multi-Layer Perceptron (MLP). Then first graph attention layer will integrate the feature vectors of node $i$ and other $n-1$  nodes and generate the latent feature vector $h_i^{\prime}$. The second graph attention layer will extract feature vector $h_i^{\prime\prime}$ further. Finally, inspired by DenseNet, the features of the preceding layers are concatenated and input into a dense layer to get the estimated $Q$ value for each action.

The graph attention mechanism in this study employs the multi-head dot-product attention approach \cite{vaswani2017attention}. This approach involves transforming the input feature of each node into query, key, and value representations by each attention head. For attention head $m$, the relationship between nodes $i$ and $j$ in the set $\mathcal{N}$ is calculated as follows:

\begin{equation}
\alpha_{ij}^{m}=\frac{exp(\Psi\cdot\mathbf{W}_q^m h_i\cdot(\mathbf{W}_k^m h_j)^T)}{\sum\nolimits_{z\in \mathcal{N}}exp(\Psi\cdot\mathbf{W}_q^m h_i\cdot(\mathbf{W}_k^m h_z)^T)},
\label{graph relation}
\end{equation}
where the weight matrices $\mathbf{W}_q^m$ and $\mathbf{W}_k^m$ in the graph attention layer are utilized to project the input feature of each node onto query and key representations respectively. The factor $\Psi$ is used to scale the dot-product of the query and key representations to ensure that the magnitude of the output is controlled. This scaling factor is typically set to the square root of the dimensionality of the query and key vectors, which is known to improve the performance of the attention mechanism.

The outputs of the $M$ attention heads for node $i$ are concatenated and then processed by the function $\sigma$, which is a one-layer MLP with ReLU non-linearities. The resulting output is given by:

\begin{equation}
h_i^{\prime} = \sigma\left(Concatenate\left[\sum\nolimits_{j\in \mathcal{N}}\alpha_{ij}^{m}\mathbf{W}_v^m h_j, \forall m\in M\right]\right),
\end{equation}
where $\mathbf{W}_v^m$ is another weight matrix, which is used to map the feature vector of each neighboring node to a new space that emphasizes different aspects of the node's relationship with the target node. The $Concatenate$ operation is used to combine the outputs of the $M$ attention heads into a single vector. The multi-head dot-product attention mechanism allows for the consideration of multiple aspects of the relationships between nodes in the graph, leading to a more nuanced representation of these relationships. The final output $h_i^{\prime}$ is a representation of node $i$ that incorporates information from its relationships with all other nodes in the graph.

\subsection{Computational Complexity}
In this subsection, we present the computational complexity of multi-head graph attention network layer. Take the first multi-head graph attention network layer as an example,
the input feature matrix is $h=\left\{h_{1},h_{2},...,h_{n}\right\}^T, h_{i}\in \mathbb{R}^{F}$, where $n$ is the number of nodes and $F$ is the number of features in each node. The dimension of input feature matrix is $(n,F)$. The weight matrix $\mathbf{W}_q$, $\mathbf{W}_k$, $\mathbf{W}_v$ is calculated from the feature matrix $h$ by multiplying three learned matrix with shape $(F,F)$. The computation complexity of this operation is $O(nF^2)$. The calculation in (\ref{graph relation}) refers to the matrix multiplication with shape $(n,F)$ and $(F,n)$, therefore, this operation has the computation complexity $O(n^2F)$. Eventually, the total complexity of multi-head graph attention network layer is $O(K(nF^2+n^2F))$, where $K$ is the number of the head. Overall, the computational complexity of the multi-head graph attention network layer scales quadratically with the number of nodes in the graph and the number of features per node, and linearly with the number of attention heads.

\section{Numerical results}
% \addtolength{\topmargin}{0.03in}
In this section, we evaluate the performance of our proposed transmission protocol optimized via the DRL algorithm using simulations. In the simulation, one GBS control center is located at the center of a circular ground area with a radius $R_{\rm G} = 300$ m, along with four fixed interfering GBSs at $(\pm 105, \pm 105, 0)$ m. The UAV swarm move in a circular aerial area with radius $R_{\rm U} = 60$ m and height $H = 300$ m. Detailed parameters are given in Table \ref{table1}.

\begin{table}[htbp]
	\centering  
	\caption{Simulation parameters}  
	\label{table1} 
	\begin{tabular}{|c|c|c|c|}
		\hline
		UAV number $N$ &5
		&Power parameter $\xi$ &1e-3\\
		\hline
		Receiving power of UAV $\tilde{P}_{i,\rm{r}}$&50 mW
		&Electronic power consumption $\tilde{P}_{i,\rm{o}}$&50 mW\\
		\hline
		Idle power of UAV $\tilde{P}_{i,\rm{idle}}$&0 mW
		&Maximum transmit power of UAV$\tilde{P}_{\rm{max}}$&23 dBm\\
		\hline
		Transmission power of GBSs $P_{\rm G}$ &43 dBm
		&Number of GBS antenna &4\\
		\hline
		Path loss exponent $\alpha_{\rm \uppercase\expandafter{\romannumeral1}}$ &$-2$
		&The broadcast frequency $f_{c}$ &2 GHz \\
		\hline
		The path loss coefficients in LoS $\eta_{\rm{LoS}}$ &$10^{-0.1}$
		&The path loss coefficients in NLoS $\eta_{\rm{NLoS}}$ &$10^{-2}$\\
		\hline
		Conversion factor of power amplifier $\kappa$ & 2.857
		&Path loss exponent $\alpha_{\rm \uppercase\expandafter{\romannumeral2}}$ &$4$\\
		\hline
		The received SINR threshold $\gamma_{\rm \uppercase\expandafter{\romannumeral1}}$, 
		$\gamma_{\rm \uppercase\expandafter{\romannumeral2}}$ &0 dBm
		&Duration of time $\tau$ &1ms\\
		\hline
		Duration of time $\tau^{(\rm \uppercase\expandafter{\romannumeral1})}$ &0.125 ms
        &Duration of time $\tau^{(\rm \uppercase\expandafter{\romannumeral2})}$ &0.875 ms\\		
		\hline
		Duration of each time slot $\Delta t$ &0.25 ms
		&Noise power $\sigma ^2$ &-90 dBm\\
		\hline
		
	\end{tabular}
\end{table}

\begin{table}[htbp]
	\centering  
	\caption{Learning parameters}  
	\label{table2} 
	\begin{tabular}{|c|c|c|c|}
		\hline
		Total episode &2000
		&The number of GNN head &8\\
		\hline
		Learning rate $\alpha$ &0.001
		&Discount rate $\gamma$ &0.98\\
            \hline
            Interpolation parameter $\beta$ &0.01
		&Replay Memory $D_i$ Capacity&2000\\
		\hline
		Minimum exploration rate $\epsilon$ &0.01
		&Minibatch size &32\\
		\hline
		Optimizer &AdamOptimizer
		&Activation function &ReLU\\
		\hline
	\end{tabular}
\end{table}
The hyperparameters for the GAT-based MADQN are listed in Table \ref{table2}. Our model has been trained over $2000$ episodes, where each episode consists of 200 rounds of C\&C message transmission. In each C\&C message transmission round, the proposed transmission protocol and DRL algorithm are simulated, where the position of the UAV swarm is assumed to be fixed during each round of C\&C transmission. The initial value of the exploration rate is set to be $0.6$ and linearly decreased to $0.01$ by multiplying 0.996 after each episode. It's noted that the proposed algorithm is fully decentralized, where each agent will train its model based on local Replay Memory. Hence, we set the capacity of Replay Memory to be $2000$ to alleviate the memory burden.

\begin{figure}[htbp]
\centerline{\includegraphics[scale=0.6]{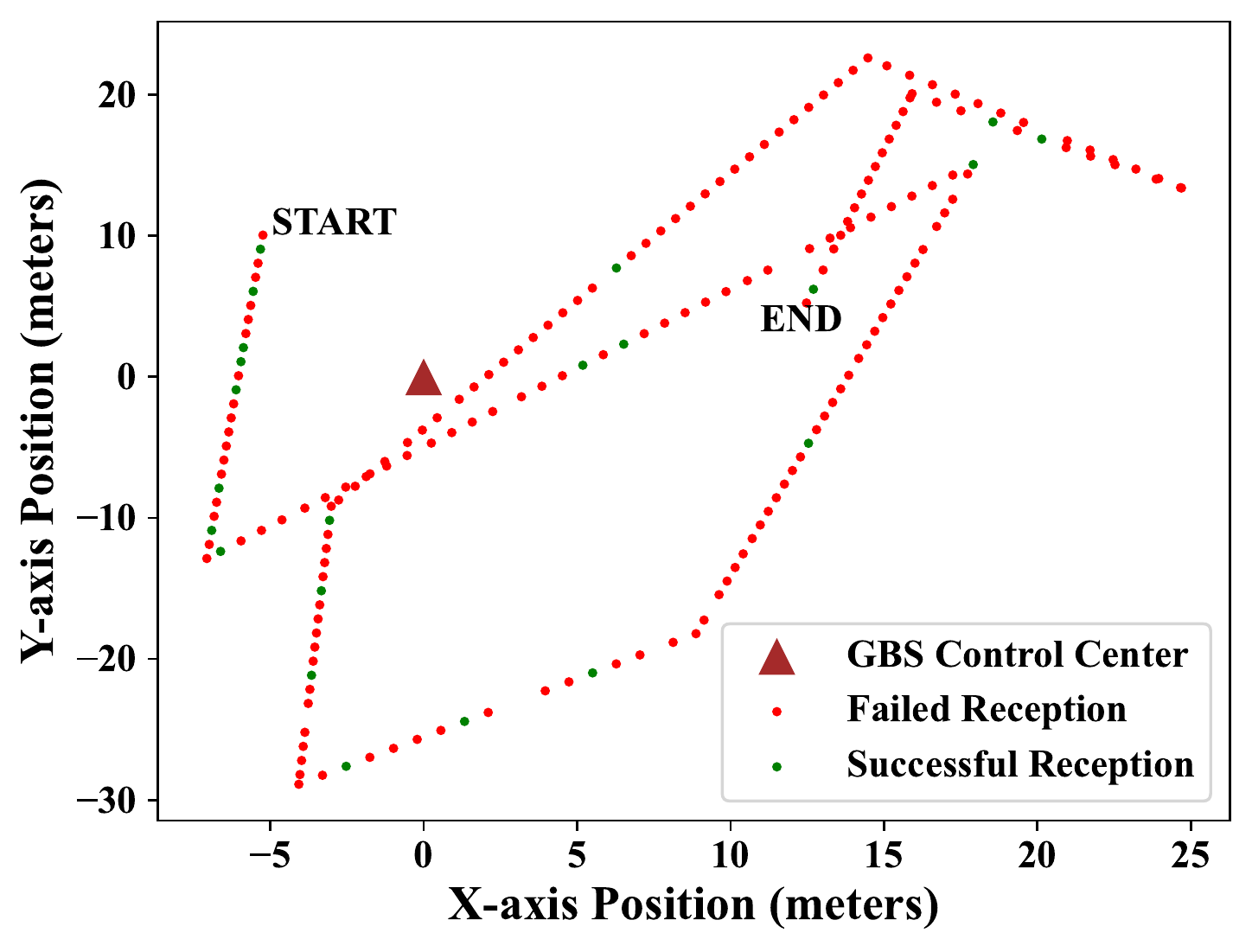}}
\caption{The moving trajectory of a UAV and corresponding common C\&C message receiving status after the broadcasting of the GBS control center.}
\label{scatter}
\end{figure}
\subsection{The state of UAV in phase I}
Fig.~\ref{scatter} plots the moving trajectory of a UAV in an episode and the corresponding common C\&C message receiving status of the UAV after the broadcasting of the GBS control center. We observe that the UAV cannot receive the message successfully in most locations due to the substantial interference from other GBSs, pathloss, and channel fading.

\begin{figure}[htbp]
    \centering
    \subfigure[]{
	\includegraphics[scale=0.58]{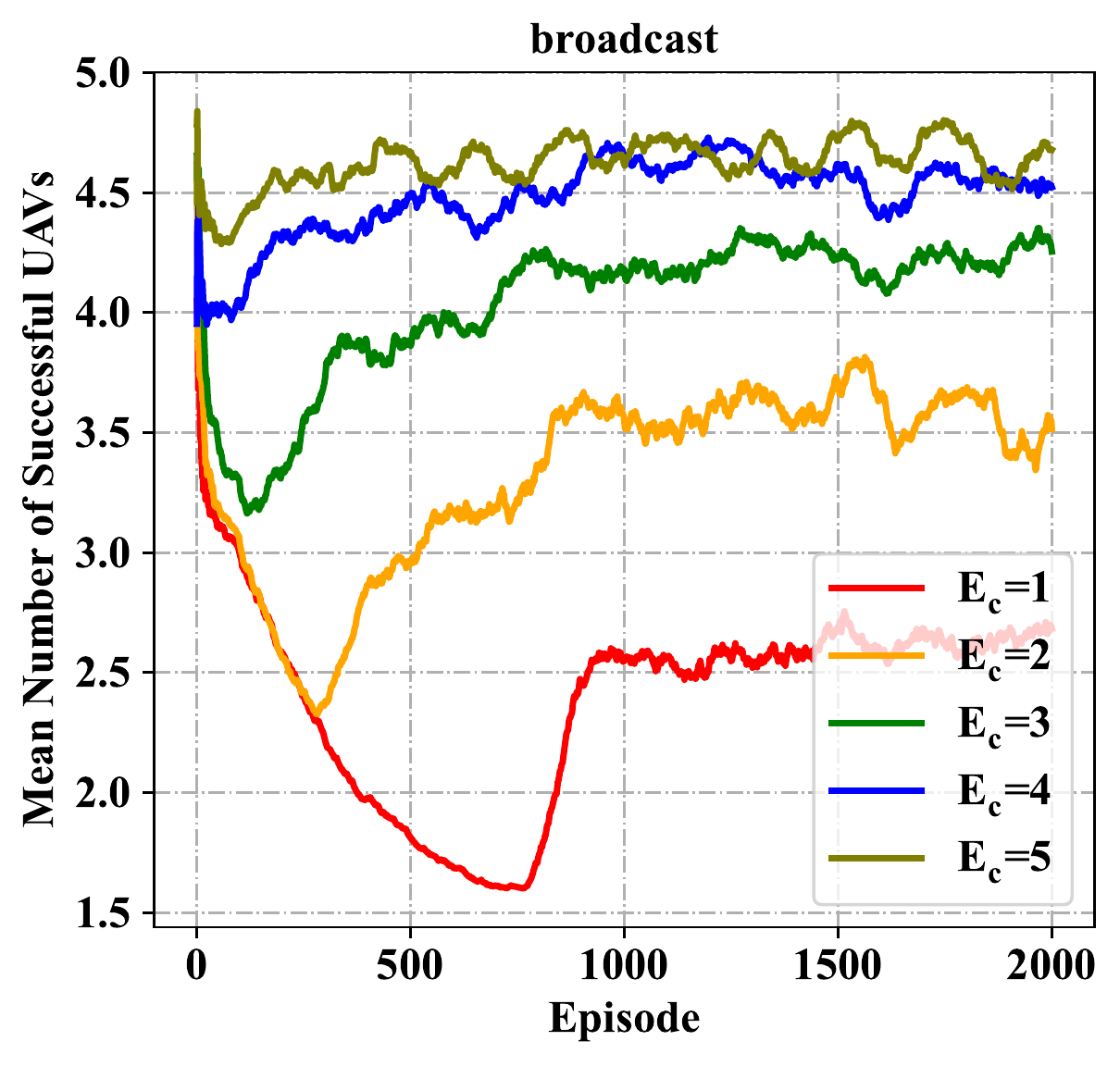}
    \label{broadcast different cons}
    }
    \subfigure[]{
        \includegraphics[scale=0.58]{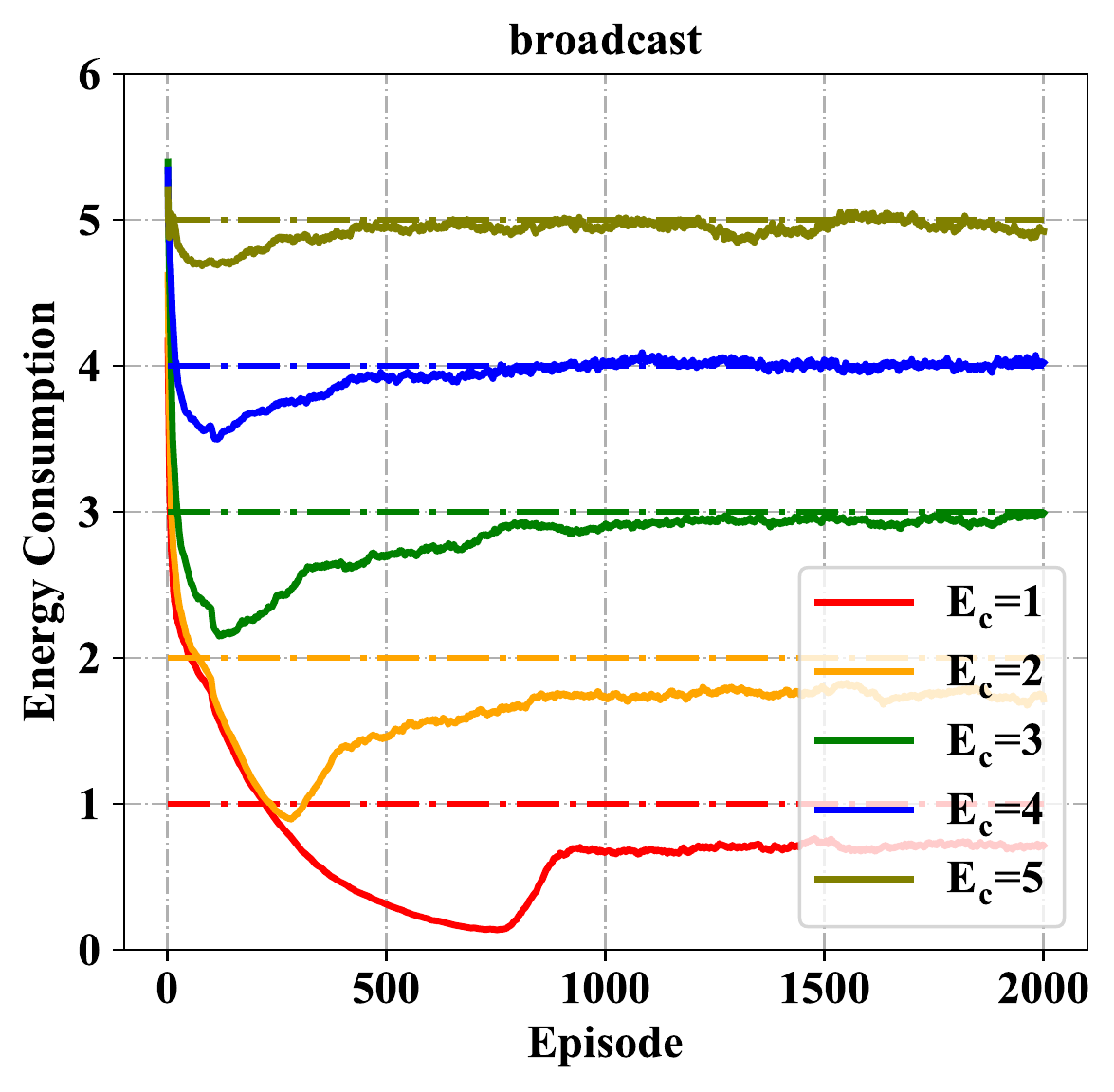}
        \label{broadcast different cons energy}
    }
    \subfigure[]{
	\includegraphics[scale=0.58]{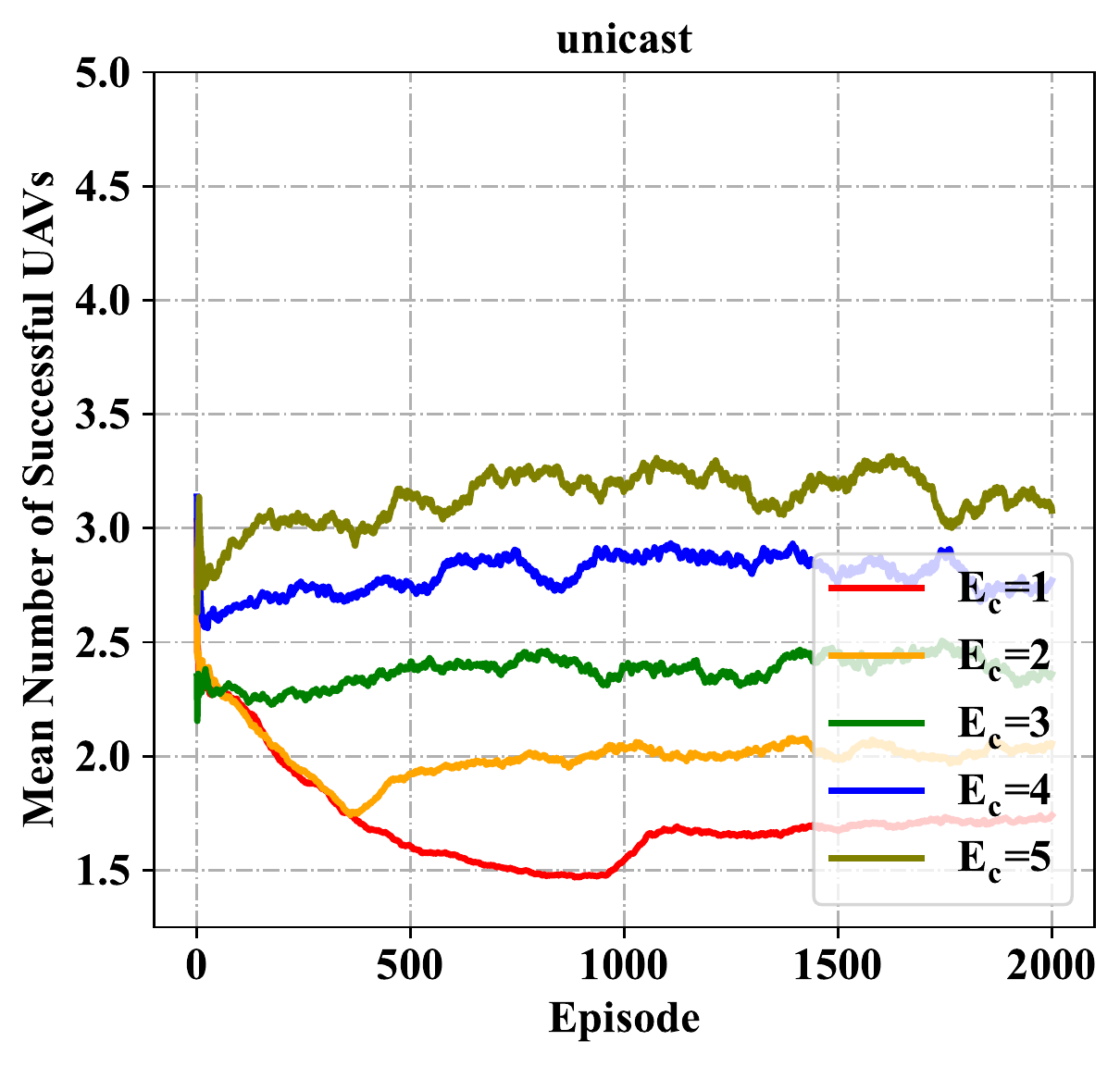}
    \label{unicast different cons}
    }
    \subfigure[]{
        \includegraphics[scale=0.58]{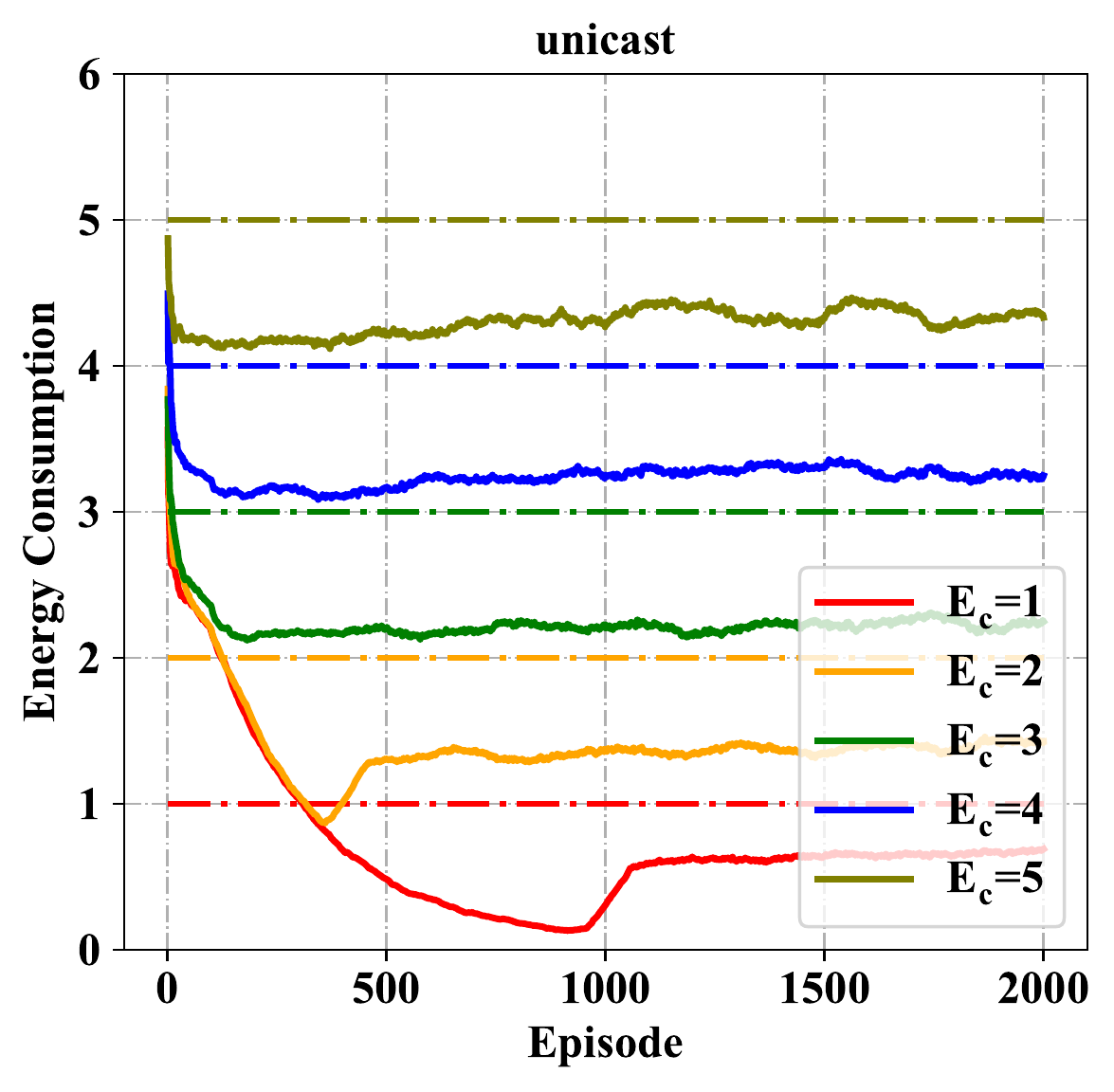}
        \label{unicast different cons energy}
    }
    \subfigure[]{
	\includegraphics[scale=0.58]{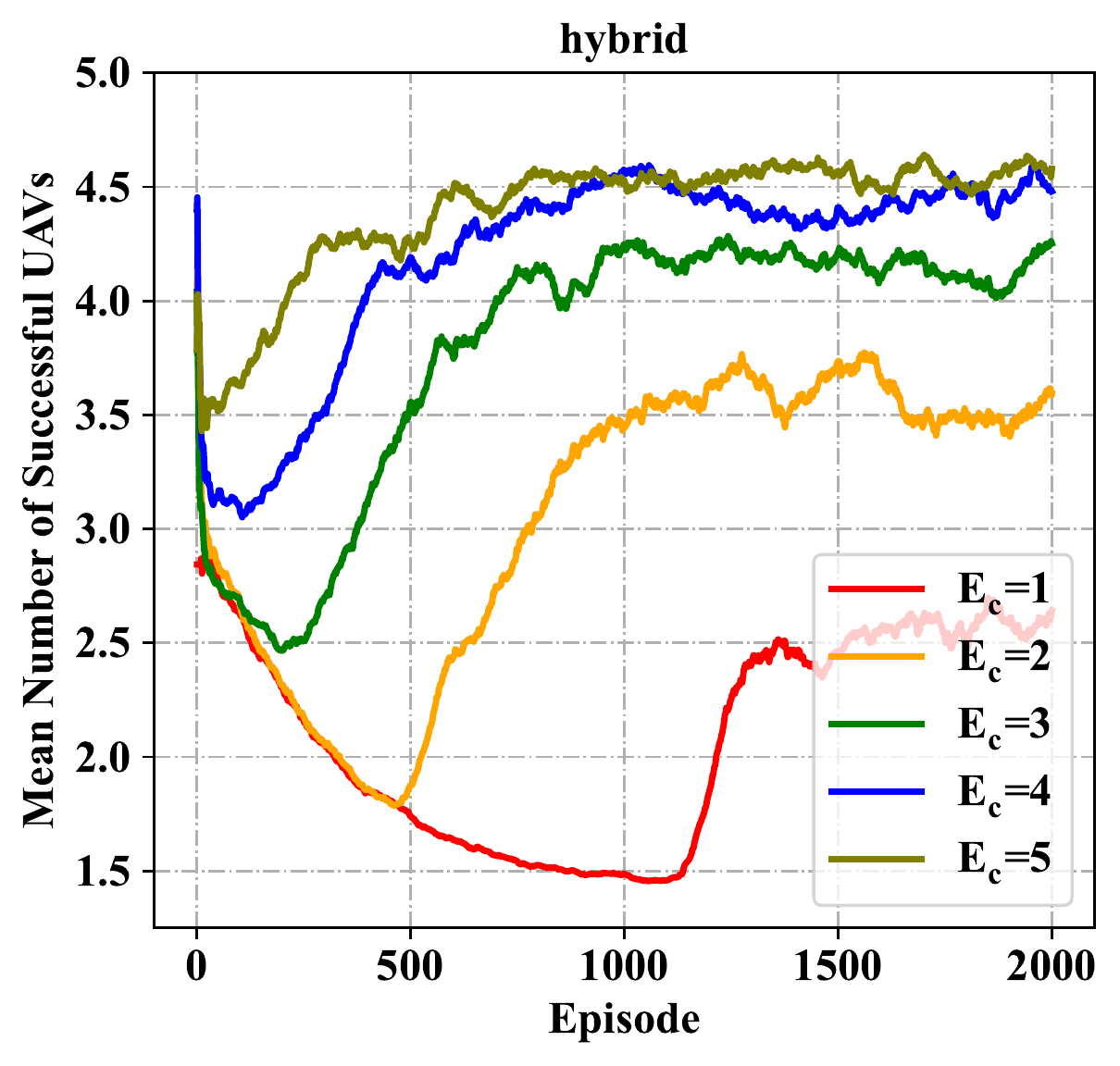}
    \label{hybrid different cons}
    }
    \subfigure[]{
        \includegraphics[scale=0.58]{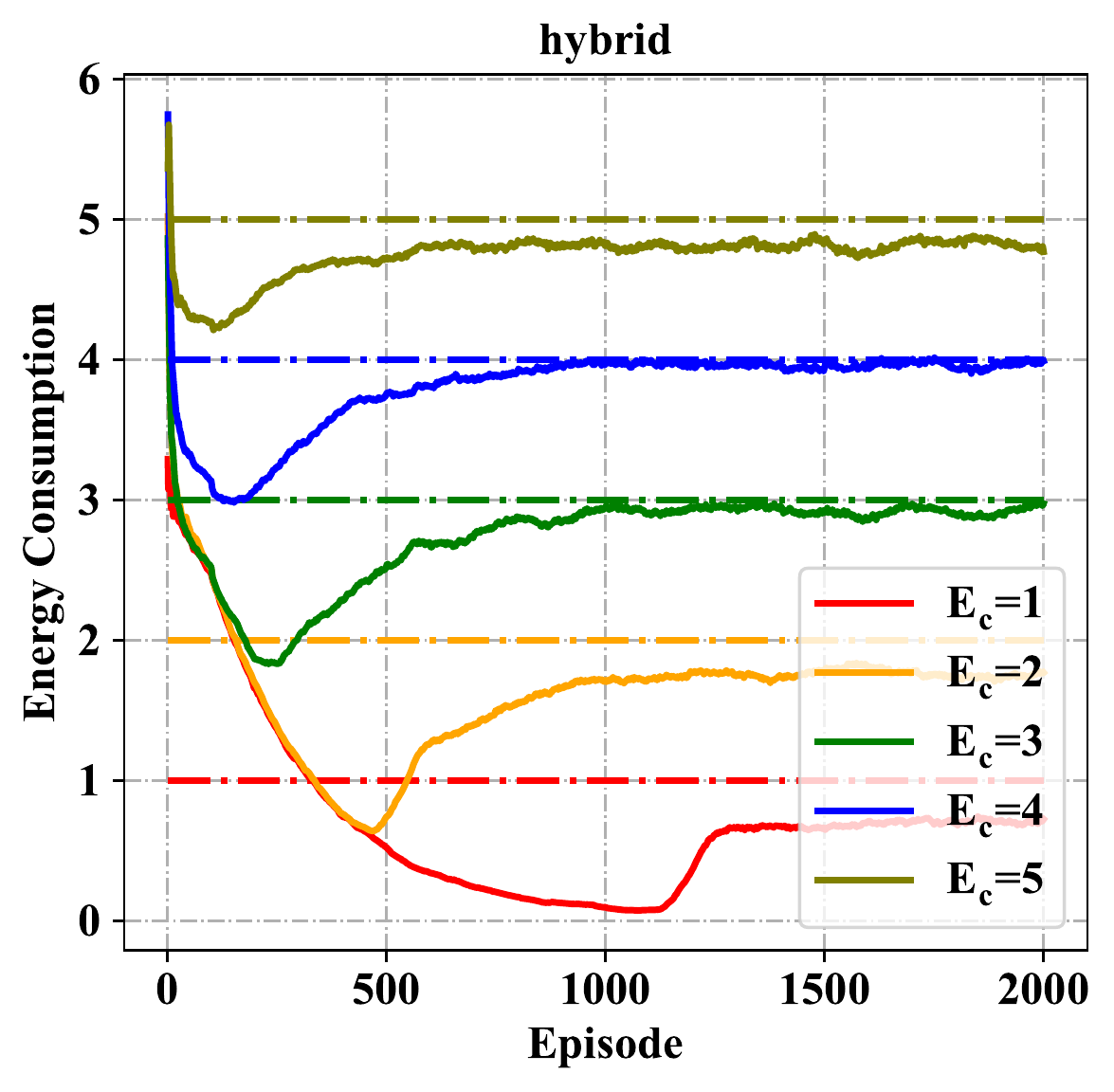}
        \label{hybrid different cons energy}
    }
    \caption{(a)(c)(e) The mean number of UAVs that successfully receive the common C\&C within latency constraints in the broadcast, unicast, and hybrid schemes under different energy constraints. (b)(d)(f) The energy consumption of the broadcast, unicast, and hybrid scheme under different energy constraints during the training process.}
    \label{three schemes}
\end{figure}
\subsection{The performance of D2D broadcast, unicast and hybrid schemes under different energy constraints}

Fig.~\ref{three schemes} presents a comprehensive performance evaluation of the D2D broadcast, unicast, and hybrid schemes under varying energy constraints during the training process. To improve readability, we normalize the energy with respect to the broadcast mode energy consumption $E_{i,\rm{b}}$. Specifically, we denote a normalized energy consumption requirement of $\rm{E_c} = 1$ as indicating that the energy consumption requirement is lower than the energy required for a one-time broadcast transmission.

During our experimental study, we have observed that the mean number of UAVs successfully receiving the C\&C message in various schemes display a consistent trend in their variation during the training process. Specifically, this trend entails a gradual decrease in the mean number of successful UAVs followed by a subsequent rise until convergence. We attribute this trend to the initial training process, where the UAV policy resulted in a high overall energy consumption that exceeded the energy constraint. However, with the implementation of the DCGA-MADQN algorithm, the UAV policy learned to consume less energy, which led to a reduction in the mean number of successful UAVs. As the energy consumption reduced to below the energy constraint, the DCGA-MADQN algorithm learned a policy that maximized the final mean number of UAVs receiving the C\&C message, resulting in a gradual increase in the mean number related curve until convergence.

We observe that our DCGA-MADQN algorithm effectively suppresses the overall energy consumption of the UAV swarm system below the predefined energy constraints during the later episodes of the training process, regardless of the energy constraint setting. Additionally, a more relaxed energy constraint setting led to an increased number of successfully receiving UAVs. This is because more energy allows for more transmission operations, whether broadcast or unicast, thereby increasing the chances of successful message transmission. However, it is shown that the mean number of successfully receiving UAVs can not be further increased after a relatively high energy constraint, this is because the policy learned by our DCGA-MADQN algorithm has explored the maximum success performance that can be achieved in the system.

\subsection{The variation of Lagrange Multiplier}
\begin{figure}[htbp]
\centerline{\includegraphics[scale=0.6]{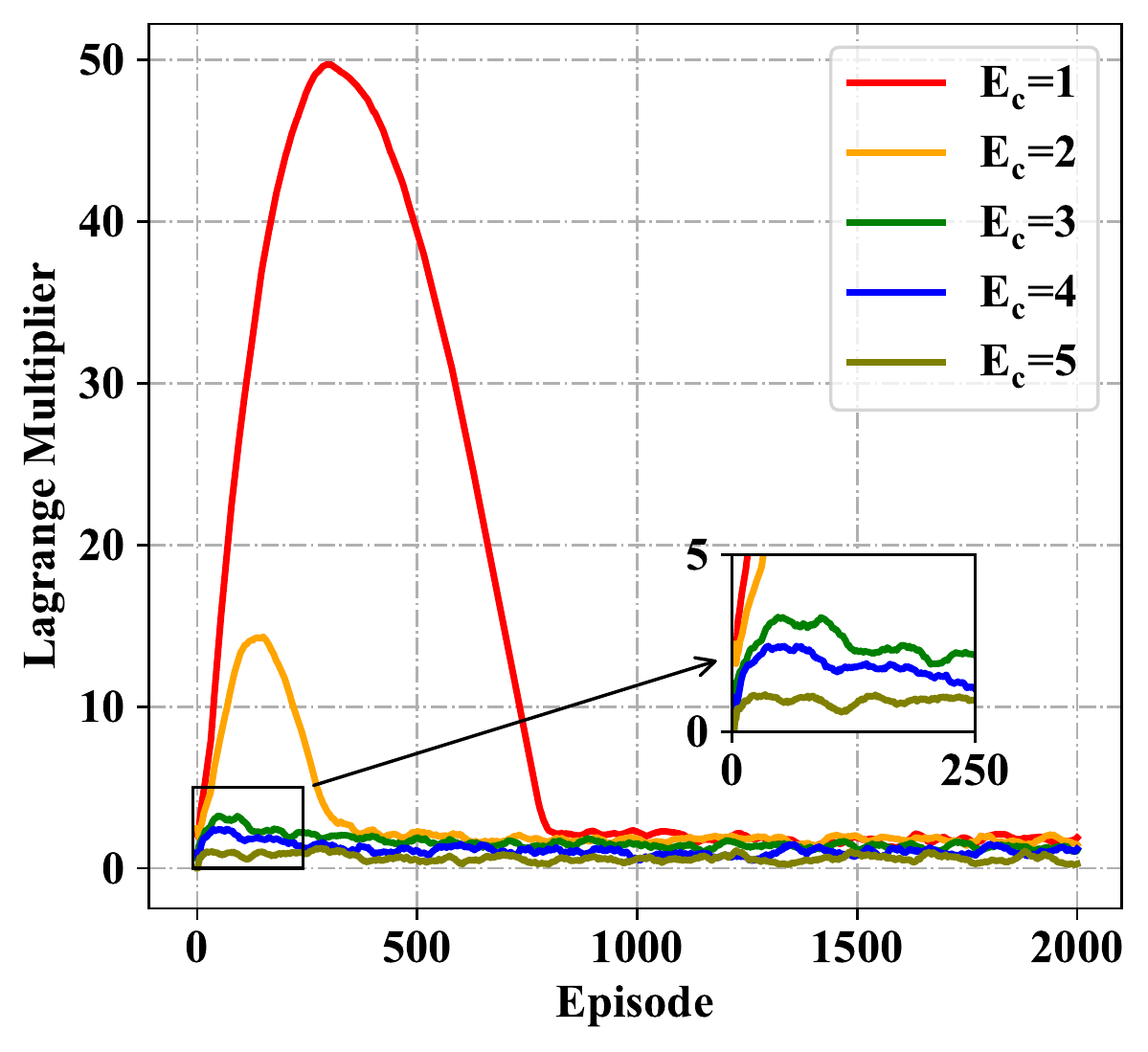}}
\caption{The variation of Lagrange Multiplier in the broadcast scheme.}
\label{lagrange multiplier broadcast}
\end{figure}

Our proposed DCGA-MADQN algorithm integrates a PID-Controller algorithm to dynamically adjust the Lagrange Multiplier, which is based on the energy consumption status of the UAV swarm and the energy constraint requirement during the training process. To illustrate the effectiveness of the proposed method, we take the broadcast scheme as an example, and plot the corresponding Lagrange Multiplier variation curve Fig.~\ref{lagrange multiplier broadcast}.

Fig.~\ref{three schemes}~\subref{broadcast different cons energy} depicts the energy consumption of the UAV swarm under different energy constraints during the training process. It is evident that at the start of the training process, the energy consumption of the UAV swarm significantly exceeds the energy constraint. The proposed PID-Controller Lagrange multiplier update algorithm effectively regulates the Lagrange multiplier $\lambda$, which gradually increases from its initial value of 0. The increase in $\lambda$ has a direct impact on the target state-action value depicted in Algorithm \ref{MADQN} and indirectly influences the learned policy of each agent. As a result of this effective control mechanism, the energy consumption curve exhibits a gradual decline. Notably, we have observed that for varying energy constraints, the rate of increase of the Lagrange multiplier $\lambda$ is variable, whereby the speed of increase is positively correlated with the disparity between energy consumption and the energy constraint. Specifically, a larger discrepancy between energy consumption and the energy constraint results in a higher rate of increase for $\lambda$.

As the energy consumption of the UAV swarm gradually approaches the energy constraint, the Lagrange multiplier reaches its maximum value. Once the energy consumption of the UAV swarm satisfies the energy constraint, and the Lagrange multiplier $\lambda$ does not increase any further. Due to the inertia of the learned policy, the energy consumption of the UAV swarm continues to decline, leading to a reduction in the value of $\lambda$. When $\lambda$ falls to a small value, it reaches an equilibrium state and fluctuates around this value. This fluctuation occurs because the controller continues to adjust the value of $\lambda$ to maintain a balance between the energy consumption and the energy constraint. 
Specifically, the controller increases the value of $\lambda$ if the energy consumption exceeds the energy constraint, and decreases the value of $\lambda$ if the energy consumption falls below the energy constraint.
Meanwhile, as the energy consumption satisfies the energy requirement, the DCGA-MADQN algorithm adapts a policy that maximizes the ultimate mean number of UAVs that receive the C\&C message. This policy optimization process elicits a slight elevation in energy consumption, albeit one that remains compliant with the energy constraint.

\begin{figure}[htbp]
    \centering
    \subfigure[]{
        \includegraphics[scale=0.6]{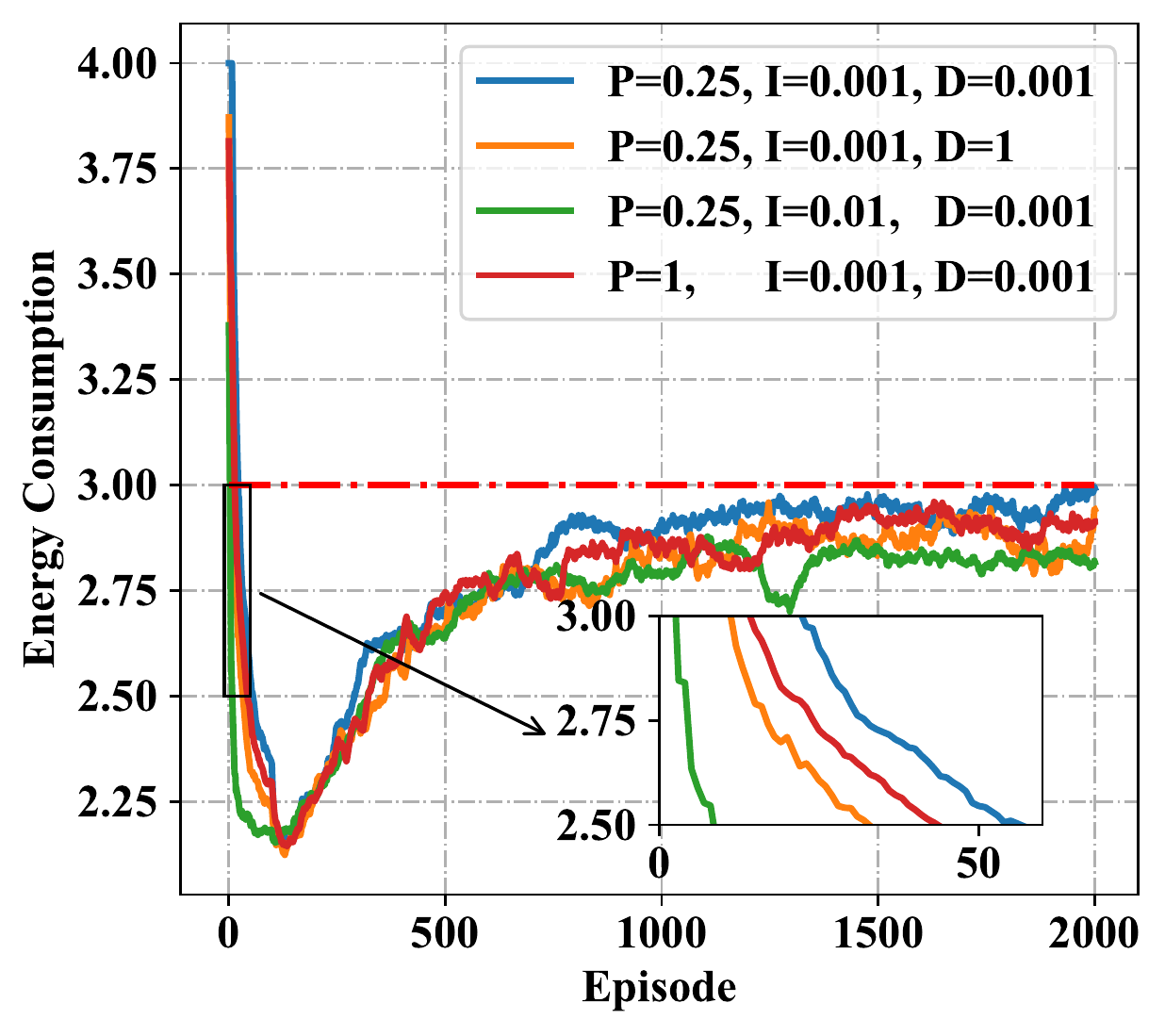}
        \label{different PID energy}
    }\hspace{20mm}
    \subfigure[]{
	\includegraphics[scale=0.6]{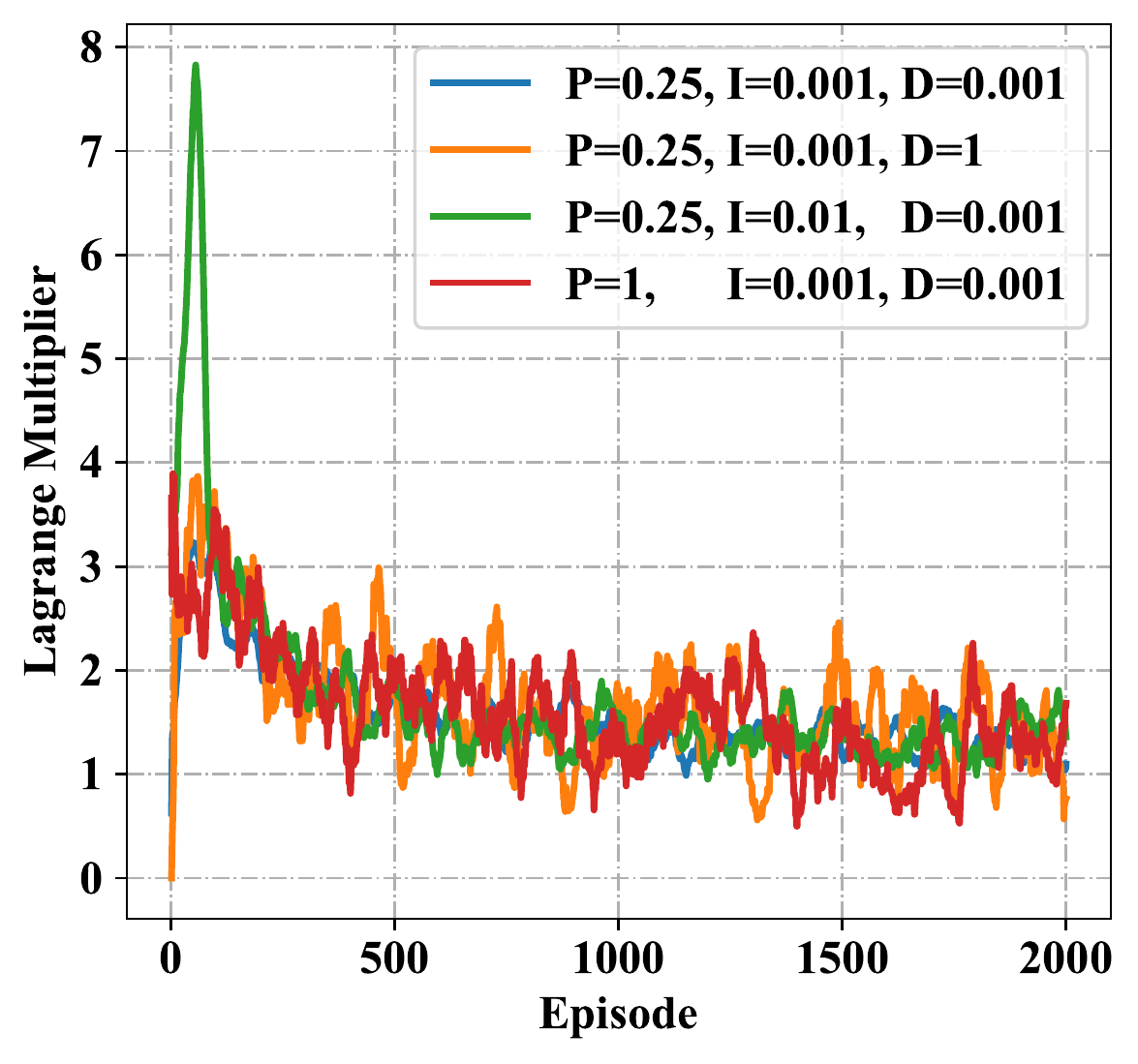}
	\label{different PID lagrange}
    }
    \subfigure[]{
	\includegraphics[scale=0.6]{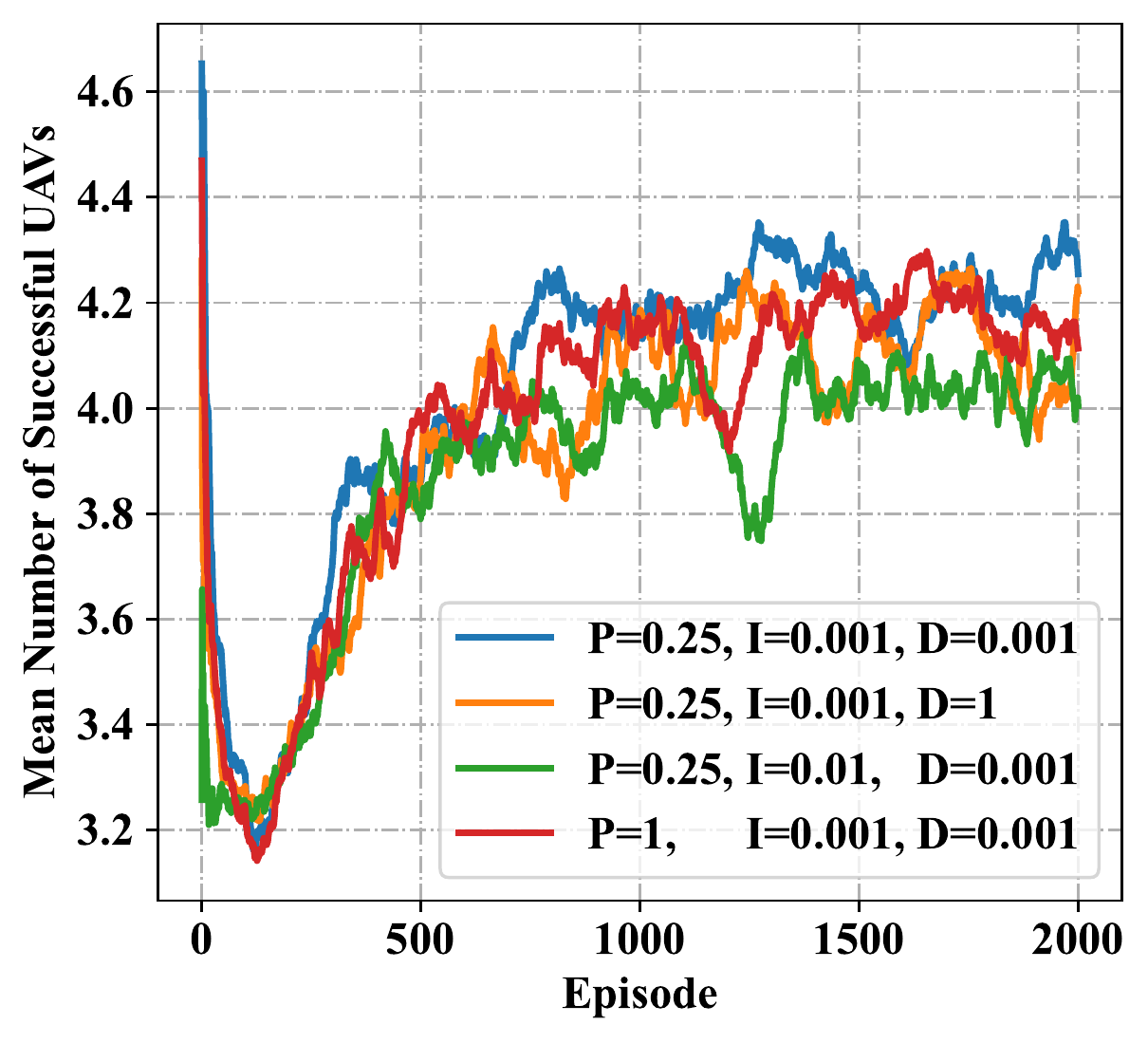}
	\label{different PID}
    }
    \caption{(a) The influence of different PID parameters for the broadcast scheme under energy constraint $\rm{E_c}=3$. (b) The variation of Lagrange Multiplier for different PID parameters (c) The influence of different PID Parameters for the average number of UAVs that successfully receive the message within latency constraint under energy constraint $\rm{E_c}=3$.}
    \label{different PID parameters}
\end{figure}
\subsection{The comparison of different PID-Controller parameters}
In Fig.~\ref{different PID parameters}, we evaluate the impact of various proportional, integral, and derivative coefficients of the PID-Controller Lagrange Multiplier algorithm on the performance of the broadcast scheme with energy constraint $\rm{E_c}=3$. 
Specifically, we establish a baseline set of parameters and generate three additional sets by adjusting a single coefficient of proportionality, integration, or differentiation at a time.
Our findings indicate that regardless of which coefficient is larger, the energy consumption curve exhibits a fast initial decrease, as demonstrated in the partial enlargement of Fig.~\ref{different PID parameters}~\subref{different PID energy}.
This is due to the larger coefficients resulting in a larger Lagrange Multiplier $\lambda$, which has a greater influence on the agents' learned policy.
Moreover, we also note that an increase in the coefficients can lead to a discrepancy between the final convergence value of the energy consumption and the energy constraint. Additionally, as depicted in Fig.~\ref{different PID parameters}~\subref{different PID}, a decrease in the final convergence value of the mean number of UAVs that successfully receive the common C\&C is observed when larger coefficients are utilized.
Furthermore, our results suggest that the integral (I) parameter has a more significant impact on the system's performance compared to the proportional (P) and derivative (D) parameters. This is attributed to the I parameter's ability to accumulate error over time, leading to a larger impact on the system's performance.

\begin{figure}[htbp]
    \centering
    \subfigure[]{
        \includegraphics[scale=0.6]{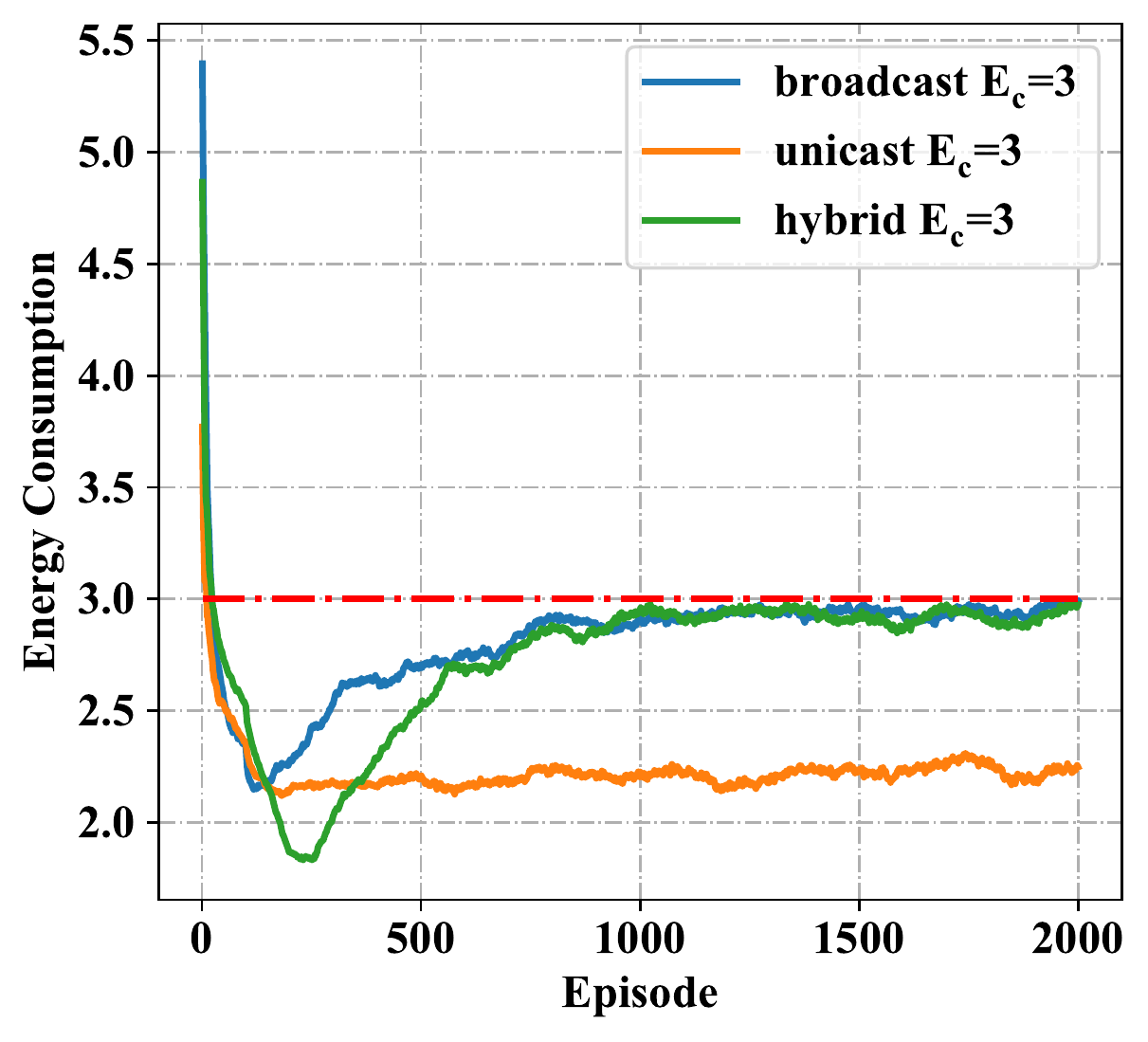}
        \label{different schemes under the same constraint energy}
    }
    \subfigure[]{
    	\includegraphics[scale=0.6]{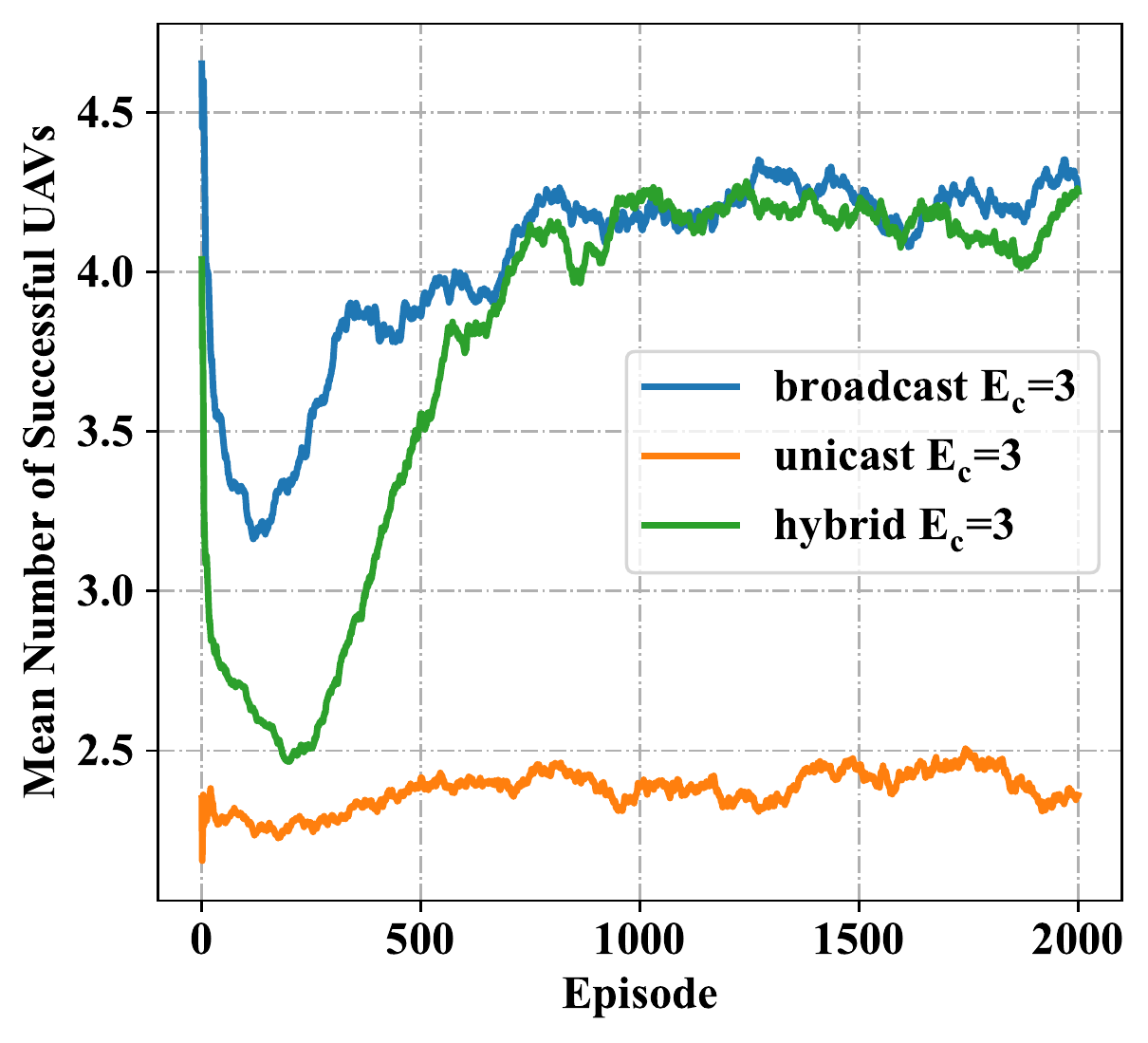}
	    \label{different schemes under the same constraint}
    }
    \caption{(a) The energy consumption of different schemes under constraint $\rm{E_c}=3$. (b) The mean number of UAVs that successfully receive the message within latency constraint in different schemes under constraint $\rm{E_c}=3$.}
    \label{same constraint}
\end{figure}

\subsection{The comparison of D2D broadcast, unicast and hybrid schemes under the same energy constraint}
We compare the energy consumption and the mean number of UAVs that successfully receive the common C\&C under the energy constraint $\rm{E_c}=3$ using the broadcast, unicast, and hybrid schemes, as shown in Fig.~\ref{same constraint}. 
We observe that during the initial training process, the hybrid scheme has a lower energy consumption compared with unicast and broadcast schemes, as shown in Fig.~\ref{same constraint}~\subref{different schemes under the same constraint energy}. 
This is because the hybrid scheme violates the energy constraint to a greater extent during the initial training process. This violation results in the DCGA-MADQN algorithm promoting the development of a policy that prioritizes the reduction of energy consumption.
Furthermore, our results also show that the mean number of UAVs that successfully receive the common C\&C in the broadcast and hybrid schemes is almost the same and significantly higher than that of the unicast scheme, as illustrated in Fig.~\ref{same constraint}~\subref{different schemes under the same constraint}. It also shows that the hybrid scheme can exploit the optimal actions that maximize the mean number of UAVs that successfully receive the common C\&C signal.

\section{Conclusion}
In this paper, we proposed a two-phase protocol to transmit the common C\&C to the UAV swarm under latency and energy constraints. To make the UAVs execute the optimal policies under the proposed three D2D schemes (unicast, broadcast, and hybrid), we design a decentralized constrained multi-agent Deep-Q-network algorithm based on the Lagrangian relaxation. Graph Attention network is utilized to learn the latent representation effectively under a highly dynamic wireless environment caused by the movement of UAVs and the change of channel state. A PID-controller method is adopted to update the Lagrange Multiplier. Simulation results show that our algorithm could effectively limit the energy consumption to the energy constraint and maximize the mean number of UAVs that successfully receive the common C\&C sent from GBS.

% \bibliographystyle{IEEEtran}
% \bibliography{IEEEabrv,mylib1}

\end{document}